\documentclass{amsart}
\newcommand{\notarxiv}[1]{}
\newcommand{\arxiv}[1]{#1}

\usepackage[normalem]{ulem}
\usepackage{multirow}
\usepackage{hhline}
\usepackage{subfigure}
\usepackage{url}
\usepackage{graphicx}

 \usepackage{indentfirst}
 \usepackage{natbib}
 \usepackage{csvsimple}
 \usepackage{float}
\makeatletter
\renewcommand{\thesubfigure}{\alph{subfigure}}
\renewcommand{\@thesubfigure}{\thesubfigure)\hskip\subfiglabelskip}
\makeatother

\newcommand{\beginsupplement}{%
        \setcounter{table}{0}
        \renewcommand{\thetable}{S\arabic{table}}%
        \setcounter{figure}{0}
        \renewcommand{\thefigure}{S\arabic{figure}}%
     }

\renewcommand{\paragraph}[1]{\smallskip \noindent\underline{#1.} }

\newcommand{\ellmax}{\ell_{\text{max}}}

\newcommand{\eat}[1]{}

\newcommand{\ptabstract}{
Bayesian Markov chain Monte Carlo explores tree space slowly, in part because it frequently returns to the same tree topology.
An alternative strategy would be to explore tree space systematically, and never return to the same topology.
In this paper, we present an efficient parallelized method to map out the high likelihood set of phylogenetic tree topologies via systematic search, which we show to be a good approximation of the high posterior set of tree topologies.
Here ``likelihood'' of a topology refers to the tree likelihood for the corresponding tree with optimized branch lengths.
We call this method ``phylogenetic topographer'' (PT).
The PT strategy is very simple: starting in a number of local topology maxima (obtained by hill-climbing from random starting points), explore out using local topology rearrangements, only continuing through topologies that are better than than some likelihood threshold below the best observed topology.
We show that the normalized topology likelihoods are a useful proxy for the Bayesian posterior probability of those topologies.
By using a non-blocking hash table keyed on unique representations of tree topologies, we avoid visiting topologies more than once across all concurrent threads exploring tree space.
We demonstrate that PT can be used directly to approximate a Bayesian consensus tree topology.
When combined with an accurate means of evaluating per-topology marginal likelihoods, PT gives an alternative procedure for obtaining Bayesian posterior distributions on phylogenetic tree topologies.
}

\newcommand{\ptkeywords}{
Bayesian phylogenetics,
consensus trees,
phylogenetic islands,
phylogenetic tree topology,
systematic search
}

\begin{document}

\arxiv{

	\title[Systematic exploration of the HLS of phylogenetic tree topologies]{Systematic exploration of the high likelihood set of phylogenetic tree topologies}

 \begin{center}

 \author[Whidden, Claywell, Fisher, Magee, Fourment, Matsen~IV]{Chris Whidden$^1$, Brian C. Claywell$^1$, Thayer Fisher$^2$, Andrew F. Magee$^3$, Mathieu Fourment$^4$,  Frederick A. Matsen~IV$^1$}

	\maketitle

 \noindent {\small \it
 $^1$Fred Hutchinson Cancer Research Center, Seattle, WA, 98109, USA\\
 $^2$Department of Biostatistics, University of Washington, Seattle, WA, 98195, USA\\
 $^3$Department of Biology, University of Washington, Seattle, WA, 98195, USA\\
 $^4$ithree institute, University of Technology Sydney, Sydney, Australia}\\
 \end{center}
 \medskip
 \noindent{\bf Corresponding author:} Frederick A. Matsen~IV, Fred Hutchinson Cancer Research Center, 1100 Fairview Ave.\ N, Mail stop M1-B514, Seattle, WA, 98109, USA; E-mail: matsen@fredhutch.org\\

}

\notarxiv{
\title{Systematic exploration of the high likelihood set of phylogenetic tree topologies}

\author{Chris Whidden$^{1}$, Brian C. Claywell$^{1}$, Thayer Fisher$^{2}$, Andrew F. Magee$^{3}$, Mathieu Fourment$^{4}$, and Frederick A. Matsen~IV$^{1,\ast}$\\[4pt]
\textit{$^{1}$~Fred Hutchinson Cancer Research Center, Seattle, WA, 98109, USA}
\\
\textit{$^{2}$~Department of Biostatistics, University of Washington, Seattle, WA, 98195, USA}
\\
\textit{$^{3}$~Department of Biology, University of Washington, Seattle, WA, 98195, USA}
\\
\textit{$^{4}$~ithree institute, University of Technology Sydney, Sydney, Australia}
\\[2pt]
\textit{{\bf Corresponding author:} Frederick A. Matsen~IV, Fred Hutchinson Cancer Research Center, 1100 Fairview Ave.\ N, Mail stop M1-B514, Seattle, WA, 98109, USA; E-mail: matsen@fredhutch.org}
}

\markboth%
{Whidden, Claywell, Fisher, Magee, Fourment, Matsen~IV}
{Systematic exploration of HLS of tree topologies}

\maketitle
}

%
%


\begin{abstract}
{\ptabstract}
{\notarxiv{\ptkeywords}}
\end{abstract}

\arxiv{
\medskip
\noindent{\bf Keywords:} \ptkeywords
\newline

\section*{Introduction}
}

Phylogenetic trees remain difficult to sample in a Bayesian framework despite decades of effort.
Some of the challenge is due to the dual nature of trees: they are composed of a discrete component, namely the tree topology, and a closely intertwined continuous aspect, such as branch lengths and substitution model parameters.
This complex model structure means that much of the recent methods developments for advanced statistical samplers in high dimensions, developed for real spaces, are difficult to apply to the phylogenetic case \citep{Dinh2017-oj}.
Thus, for exploring phylogenetic trees, we are left with uninformed random modification proposals.
Although these random methods can be carefully tuned to modify trees ``the right amount'' to get an appropriate acceptance rate \citep{Lakner2008-wu}, this can lead to timid Metropolis-Hastings moves and consequent slow progress through tree space.
Such slow progress using random modifications is a natural consequence of the fact that there are many tree topologies, yet for informative data sets the posterior is concentrated on a small fraction of them.

Indeed, there are many more ways to make a reasonable tree topology quite bad than there are to make it better or maintain the same level of optimality.
As a simple example, in the dataset DS1 (detailed below) the maximum a posteriori (MAP) topology has 2256 neighboring topologies reachable via a single standard subtree prune-and-regraft (SPR) operation.
Only 36 of these topologies are in the MrBayes 95\% credible set, so the vast majority of proposed random modifications of the MAP topology fall outside the credible set.
These issues are inherent in having a very large number of possible topologies, yet not having an efficient means of sifting through the bad topologies \emph{a priori} to find the good ones.
To make matters worse, multiple modes separated by low posterior topologies are frequently present in real data \citep{Whidden2015-eq}.

MCMC is inefficient for the purpose of visiting every topology in the credible set.
Typical credible sets have hundreds of topologies (or more) that contribute significantly in aggregate to the posterior, even if each one has significantly less posterior mass than the MAP topology.
To find all of these topologies, we need to visit the MAP topology many times.
For example, in one dataset (DS1, detailed below) the MAP topology has 7.36\% of the posterior probability while the least likely topology in the 95\% credible set has 0.00299\% of the posterior probability.
The MAP topology must be sampled 2465 times more often than non-trivial unlikely topologies; adding nearly-MAP topologies into consideration further emphasizes the point.

The power to quickly identify a set $S$ of tree topologies that neatly contain the credible set would enable new efficient strategies for Bayesian phylogenetics.
A simple MCMC approach to use $S$ would be to make bold topology proposals within $S$, which will have a high probability of acceptance, and maintain the rest of the standard MCMC machinery for continuous parameters.
An alternate approach would be to bypass MCMC altogether and evaluate continuous aspects of tree models separately using methods dedicated to the task of per-topology marginal likelihood estimation \citep{dubious}.

In this paper, we develop an efficient parallelized method to find a collection of high-likelihood topologies, which we call \emph{phylogenetic topographer} or PT for short.
Here ``likelihood'' of a topology refers to the tree likelihood for the corresponding tree with optimized branch lengths.
The PT strategy is very simple: starting in a number of local tree topology maxima (obtained by hill-climbing from random starting points), explore out using local tree rearrangements, only exploring topologies that are better than than some threshold below the best observed topology.
By using a non-blocking hash table keyed on unique string representations of tree topologies, we avoid visiting topologies more than once across all concurrent threads exploring tree space.
Here we focus on the core development of PT and consequent observations about topology likelihoods rather than applications, but as a first application show good concordance between the results of PT and MrBayes on standard test data sets.
We call the set of tree topologies explored by PT the high likelihood set, or HLS, in analogy to the high posterior density set (HPDS).

\section*{Materials \& Methods}

\begin{figure}
	\centering
	\includegraphics[width=\textwidth]{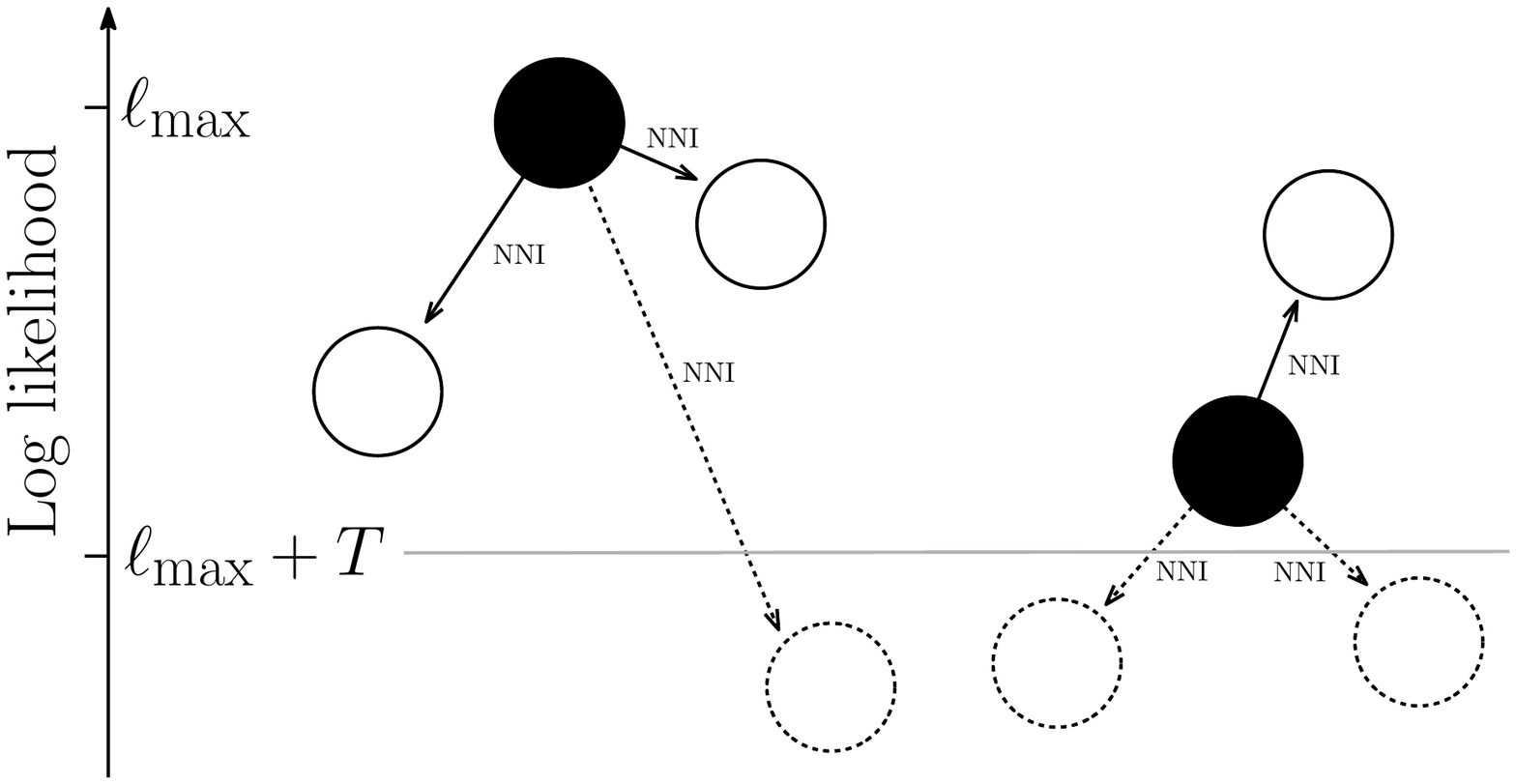}
	\caption{A high level overview of Phylogenetic Topographer.
	Starting with a set of high likelihood topologies (filled black circles), PT tests their NNI neighbors.
	Given a negative threshold $T$, the search explores all topologies above the threshold $\ellmax+T$ (hollow solid circles) while topologies below the threshold are rejected (hollow dashed circles).
		}
		\label{fig:overview}
\end{figure}

\subsection*{Methods Overview}
The goal of Phylogenetic Topographer (PT) is to identify the collection of phylogenetic tree topologies with a likelihood above a given threshold from the maximum, which we call a high likelihood set (HLS, Fig.~\ref{fig:overview}).
It does so via direct and systematic exploration, which we implemented in C++ (\url{https://github.com/matsengrp/pt}) using the libcuckoo~\citep{li2014algorithmic} nonblocking distributed hash table and libpll~\citep{libpll} Phylogenetic Likelihood Library.
PT begins with initial set of topologies and add them to a queue.
PT then visits every tree topology obtainable via NNI operations from the queued topologies.
Visiting a topology consists of optimizing branch lengths (and optionally model parameters) to maximize its log-likelihood.
Topologies within a user-specified negative threshold $T$ from the highest found log-likelihood $\ellmax$ are added to the queue and other topologies are rejected.

We tested PT on 9 empirical datasets that have become standard for evaluating MCMC methods and compare each HLS to MrBayes credible sets.
We also compared split frequencies estimated from the HLS and credible sets and visualized exploration of the credible sets as a graph.

\subsection*{Maximum Likelihood Estimation of Local Maxima}
\label{sec:methods:raxml}
To generate the initial set of PT topologies for further exploration, we inferred maximum likelihood topologies using RAxML~\citep{stamatakis2014raxml}.
If there are multiple local modes, also known as phylogenetic islands \citep{Maddison1991-tc}, then we want to include topologies from each mode in the set of starting topologies.
Although RAxML does randomize the addition order when building an initial topology via stepwise addition of taxa to maximize the parsimony score, this does not always find multiple modes in tree space.

To find a diverse set of topologies covering multiple phylogenetic islands, we ran RAxML 200 times for each dataset with random starting topologies generated using a birth-death process.
These topologies were generated using DendroPy~\citep{sukumaran2010dendropy} with birth and death rates both set to 1.0.
We used the \citet{jukes1969evolution} substitution model with the gamma model of rate variation across sites to optimize the random starting topologies (RAxML commands: \texttt{--JC69 -m GTRGAMMA}).
We compared the results of running PT with 200 such optimized random starting points and one single maximum likelihood starting point generated with stepwise addition.

\subsection*{Phylogenetic Topographer}
\label{sec:methods:pt}
The input to PT is a phylogenetic tree or set of trees in Newick format, a RAxML info file containing the desired phylogenetic model parameters, and a desired negative log-likelihood threshold offset $T$.
PT searches for every tree topology with a maximum log-likelihood within $T$ log units of the highest log-likelihood $\ellmax$ encountered.
Again, we used the Jukes-Cantor plus gamma substitution model in this paper and single starting trees or sets of starting trees as described above.
However, more complex models such as the general time-reversible model~\citep{lanave1984new,tavare1986some} are supported by PT, as are any other models supported by libpll.

The key idea of PT is to visit tree topologies in the general neighborhood of the starting trees that have a high maximum likelihood, specifically at least $\ellmax + T$.
The set of neighboring topologies are defined by the nearest-neighbor interchange (NNI) operation.
NNI operations are a subset of SPR operations and are also commonly used in phylogenetic methods.
A topology with $n$ leaves has $O(n)$ NNI neighbors as opposed to $O(n^2)$ SPR neighbors so NNI operations provide a quicker but more local search.
We compensate for the local nature of NNI operations by having many starting points.

We maintain three distributed datastructures.
The \emph{todo} queue stores the possible high likelihood topologies which have not yet been visited.
Partial optimization determines whether topologies go onto the todo queue.
Every topology in the todo queue will eventually be visited, i.e.\ will have its branch lengths fully optimized.
The \emph{visited} hash table stores every topology for which branch lengths have been fully optimized.
The \emph{good} hash table stores the high likelihood topologies.
Tree topologies are stored using an SDLNewick string representation~\citep{whidden2018efficiently}, a left/right sorted variant of the Newick string format that ensures the same string representation for identical tree topologies.

The todo queue is initialized with a set of input topologies; PT performs the following steps for each topology in the todo queue until it is exhausted:
\begin{enumerate}
 	\item Add the topology to the visited hash table and remove it from the todo queue.
	\item Optimize the branch lengths of the corresponding tree to maximize its likelihood.
	\item If the tree likelihood is less than $\ellmax + T$ then discard the tree.
	\item Otherwise, add the topology with its likelihood to the good hash table.
	\item Update $\ellmax$ if the tree likelihood is greater than it.
	\item For each unvisited NNI neighbor of the tree: \begin{enumerate}
		\item Optimize the branch length of the new tree branch and optionally other tree branches.
		\item Discard the neighbor if its partially-optimized likelihood is less than $\ellmax + T$.
		\item Otherwise, add the neighbor tree topology to the todo queue.
\end{enumerate}
\end{enumerate}

By default, PT applies single branch optimization when proposing neighboring trees, that is, PT only optimizes the length of the new tree branch.
PT can optionally optimize branches at a certain radius around the new branch with the \verb=--radius= $r$ option.
As defined by libpll, this radius optimization first selects an arbitrary neighboring node of the new branch.
Edges at most distance $r$ from the selected neighbor are optimized.
For example $r=0$ is single branch optimization, $r=1$ optimizes the new branch and two of its neighbors, and so on.
Finally, $r=-1$ specifies full branch optimization when proposing a neighbor tree.
Note that full branch optimization is always used before judging if a tree topology belongs in the good hash table and thus for the final reported log likelihood values.

We implemented a multithreaded \emph{wanderer} framework for PT to explore the high likelihood set in parallel.
A central \emph{authority thread} controls the assignment of topologies from the todo queue to wanderer threads.
When a wanderer thread exhausts its local region of tree space, the authority \emph{steals} topologies from the neighborhood of another wanderer to ensure that every wanderer thread is constantly working.

PT can either use the inferred maximum likelihood model parameters from RAxML or spend significantly more time optimizing model parameters to maximize the likelihood of each visited topology.
In our simplified case, for example, model optimization means optimizing the shape of gamma distributed rate heterogeneity.
PT can also optionally incorporate an exponential branch length prior and optimize MAP topologies instead of maximum likelihood topologies.

\subsection*{Data and Run-time Parameters}
\label{sec:methods:data}

\begin{table}
\centering
\small
\caption{The data sets used in this study, DS1-8 and DS10 of eukaryote sequences.
N~= number of species; Cols = number of nucleotides; rDNA = ribosomal DNA; rRNA = ribosomal RNA; mtDNA = mitochondial DNA; COII = cytochrome oxidase subunit II; NPC = Nuclear protein coding.
}
	\csvreader[tabular=lrrllr,
			head to column names = false,
			table head=\hline Data & N & Cols & Type of data & Study & Treebase \\\hline,
			late after line=\\,
			late after last line=\\\hline]%
		{"data/dataset_table.csv"}%
		{}%
		{\csvcoli & \csvcolii & \csvcoliii & \csvcoliv & \citet{\csvcolv} & \csvcolvi}%
\label{table:datahohna}
\end{table}

We investigated the ability of PT to find high posterior density sets by applying PT and MrBayes 3.2~\citep{ronquist2011draft} to 9 empirical data sets.
These data sets, which we call DS1-DS8 and DS10, are standard data sets for evaluating MCMC methods~\citep{Lakner2008-wu,hohna2012guided,larget2013estimation,Whidden2015-eq}.
The data sets consist of sequences from 27 to 67 eukaryote species (Table~\ref{table:datahohna}), and are fully described in \citet{Lakner2008-wu}.

Single-threaded tests were run on Intel Xeon E3-1270 Processors running Ubuntu 14.04 with 32GB of RAM.
Multithreading tests were run on Intel Xeon E5-2697 Processors running Ubuntu 14.04 with 768 GB of RAM.

To determine the high posterior density sets in these datasets we computed large ``golden run'' posterior samples for each data set using MrBayes.
In other words, we sampled far more topologies than typically used for such analyses: for each data set, 10 single-chain MrBayes replicates were run for one billion iterations and sampled every 100 iterations.
We used the \citet{jukes1969evolution} substitution model with the gamma model of rate variation across sites (MrBayes version 3.2.5 commands: \texttt{lset nst=1 rates=gamma; prset statefreqpr=fixed(equal)}).
We used a uniform prior on topologies and an unconstrained Exponential(10) prior on branch lengths.
These replicates were not Metropolis-coupled.
We discarded the first 25\% of samples as ``burn-in'' for a total of 7.5 million posterior samples per data set, assuming the long burn-in period implied stationarity.
Following~\citet{hohna2012guided} and~\citet{Whidden2015-eq}, we assumed these runs accurately estimated split frequency distributions because of the extreme length of the Markov chains in comparison to our data size.
The estimated split frequency error was below 0.015\% for each of our data sets, suggesting that the various golden runs sampled the same split frequencies.
Moreover, commonly applied diagnostics implemented in the MrBayes \texttt{sumt} and \texttt{sump} tools satisfied common thresholds, including having a standard error of log-likelihood at most 0.8, maximum standard deviation of split frequencies at most 0.006, maximum Gelman-Rubin split PSRF values of 1.000, and the effective sample size for the treelength parameter exceeding 470,000.
We use the 95\% credible set of tree topologies from the golden runs as our assumed high posterior density set.
Despite the length of these runs, we cannot assume that the golden runs have accurately estimated the posterior probability of \emph{all} topologies for all data sets.
Following previous work~\citep{Whidden2015-eq} we can, however, reasonably assume that the golden runs accurately estimate the posterior probability of high probability topologies.

In one set of comparisons considering the proportion of the high posterior density set found by PT runs, we normalized the covered density to exclude the topologies only sampled once out of all 10 golden runs; we call this the ``nonsingleton covered posterior probability.''
For example, suppose PT found topologies covering 60\% of the posterior probability, only 80\% of the posterior probability of the 95\% credible set topologies were from topologies sampled twice or more, and none of the topologies found by PT were singletons.
Then the nonsingleton covered posterior probability would be $0.6 / 0.8 = 0.75$ for such a run.
This allowed us to better evaluate an HLS for datasets where even golden runs were insufficient to fully estimate the probability of the entire 95\% credible set.
Note that we did not evaluate our methods on the datasets DS9 and DS11 from previous studies because of the difficulty in obtaining good posterior probability estimates of tree topologies for comparison on these datasets, even through extraordinarily long MrBayes golden runs.
Indeed, in previous studies~\citep{Whidden2015-eq} no tree topology in these datasets was ever sampled twice even in long MrBayes golden runs.

\subsection*{Split Frequency Comparison}
\label{sec:methods:split_frequency}

Researchers are often interested in summary measures such as consensus trees rather than the marginal probabilities of individual topologies.
Towards this end we computed PT split frequency estimates for an HLS as the sum of the probability-normalized maximized likelihood of each tree topology containing a given split (here by \emph{maximized likelihood} we mean the likelihood of the topology with optimized branch lengths).
We compared these estimates to the split frequencies inferred by MrBayes probability estimates using the same amount of computational time.
Comparisons were both quantitative by computing the root mean square deviation (RMSD) from the assumed golden run ground truth split frequencies and qualitative via scatter plots and by comparing the inferred majority rule consensus trees.
The root mean square deviation is similar to the average standard deviation of split frequencies (ASDSF) convergence diagnostic commonly used as a stopping criterion when comparing independent Markov chain Monte Carlo chains.
As such, we consider a 0.05 RMSD to be a good estimate of the split frequency distribution.
It is difficult to estimate the frequency of rare splits so, following common methods of calculating ASDSF values, we only compared the RMSD on nontrivial splits with at least 10\% frequency in the golden runs.

\subsection*{SPR Tree Space Graphs}
\label{sec:methods:spr_graphs}

We visualized exploration of the tree space of Bayesian phylogenetic posteriors using subtree prune-and-regraft (SPR) graphs constructed by the software package sprspace~\citep{Whidden2015-eq}.
This software constructs a graph in which the vertices represent the phylogenetic tree topologies of the 95\% credible set or, if that is too large, the 4096 most probable tree topologies.
The edges of these SPR graphs connect topologies which can be obtained from each other with one subtree prune-and-regraft operation.
Note that NNI operations are a subset of SPR operations so these graphs connect topologies that may not be easy to traverse by PT.
The graphs were visualized with the open-source Cytoscape platform~\citep{shannon2003cytoscape}.
We used a force-directed graph layout which repels nodes but pulls edges.
Node sizes (area) were scaled in proportion to their posterior probability.
The largest node was the most probable tree topology.
We colored nodes on a dark blue to light blue scale (grey in the print version) with increasing PT thresholds.
The darkest nodes were found with the smallest threshold.
Yellow nodes were not found by PT with the tested thresholds.

\section*{Results}

\subsection*{PT Can Quickly Find High Posterior Density Sets}

\begin{figure}
	\centering
	\subfigure[\label{fig:coverage}]{\includegraphics[width=0.9\textwidth]{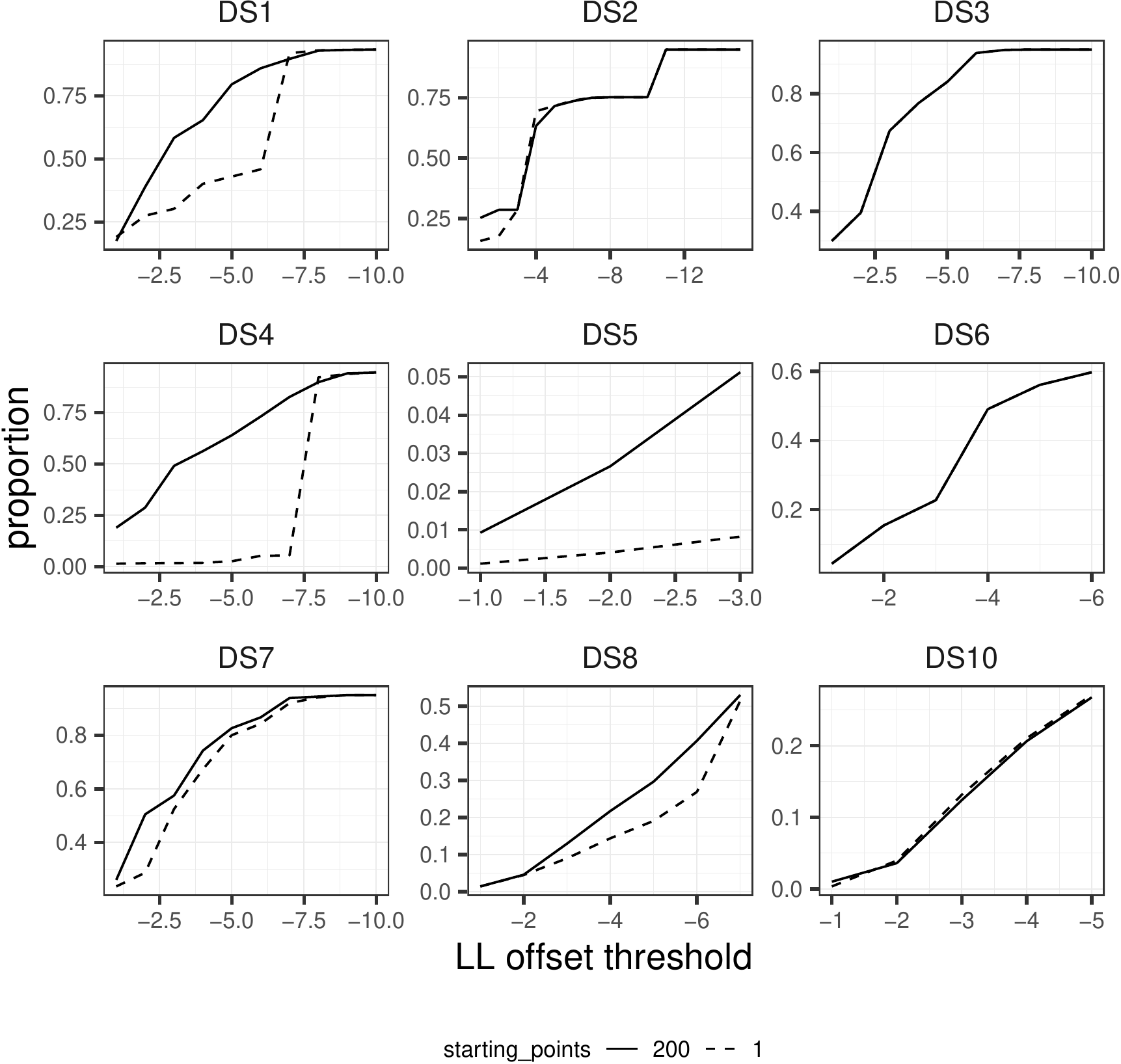}}
	\subfigure[\label{fig:coverage_time_subset}]{\includegraphics[width=0.9\textwidth]{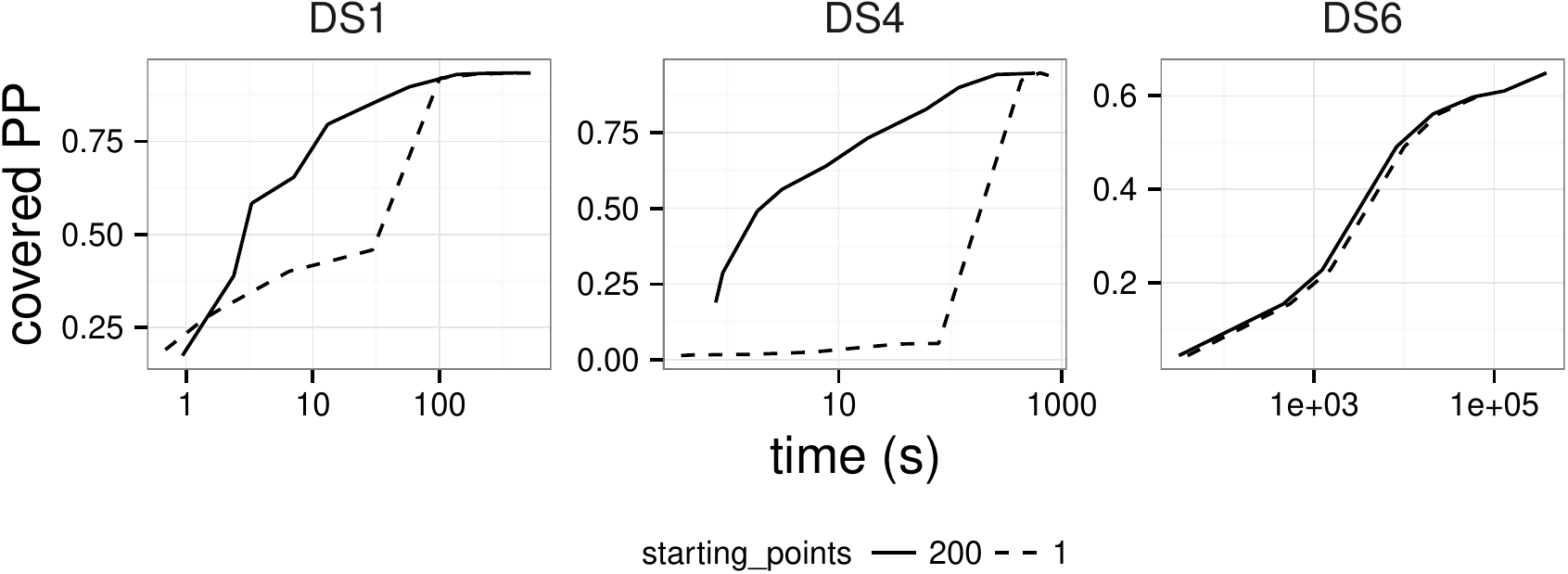}}
	\caption{
    (a) The cumulative posterior probability of HLSs explored by PT at different log-likelihood thresholds.
		(b) The covered PP (from MrBayes) with increasing time on a subset of datasets.}
    \label{fig:coverage_overall}
\end{figure}

PT finds Bayesian high posterior sets of topologies by evaluating log likelihoods with a fixed gamma rate distribution (Fig.~\ref{fig:coverage_overall}).
As expected, PT finds more of the MrBayes 95\% credible set as the search threshold increases (Fig.~\ref{fig:coverage}).
For datasets with a smaller credible set (DS1,DS3-DS4 and DS7), a threshold of -10 log-likelihood units was sufficient to capture the entire 95\% credible set.
Dataset DS2 has a few topologies with a particularly high relative likelihood which necessitated expanding the threshold to -15 to capture the entire 95\% credible set.
For datasets with a moderately sized credible set (DS6, DS8), a threshold of -6 or -7 log-likelihood units captured topologies with the majority of the credible set posterior probability.
These datasets exhausted the available 32GB memory and failed with larger thresholds.
In datasets with a wide, flat posterior (DS5, DS10), and thresholds of -3 and -5, PT captured only the small proportion of the credible set posterior probability focused on topologies with a relatively high posterior probability.

Using multiple starting points is important when exploring peaky datasets with multiple modes.
Previous work~\citep{Whidden2015-eq} identified DS1, DS4, DS6, and DS10 as having multiple modes.
Using 200 starting points captured a much larger proportion of the credible set on DS1 and DS4 at the same threshold.
When using a single starting point these datasets showed a characteristic ``bump'' in the covered probability at the threshold which allowed PT to find a second mode.
However, PT showed no or little difference when exploring DS6 and DS10 with one or multiple starting points, nor was this bump evident.
Multiple starting points also aided the exploration of the flat dataset DS5.

\begin{figure}[H]
	\centering
	\includegraphics[width=\textwidth]{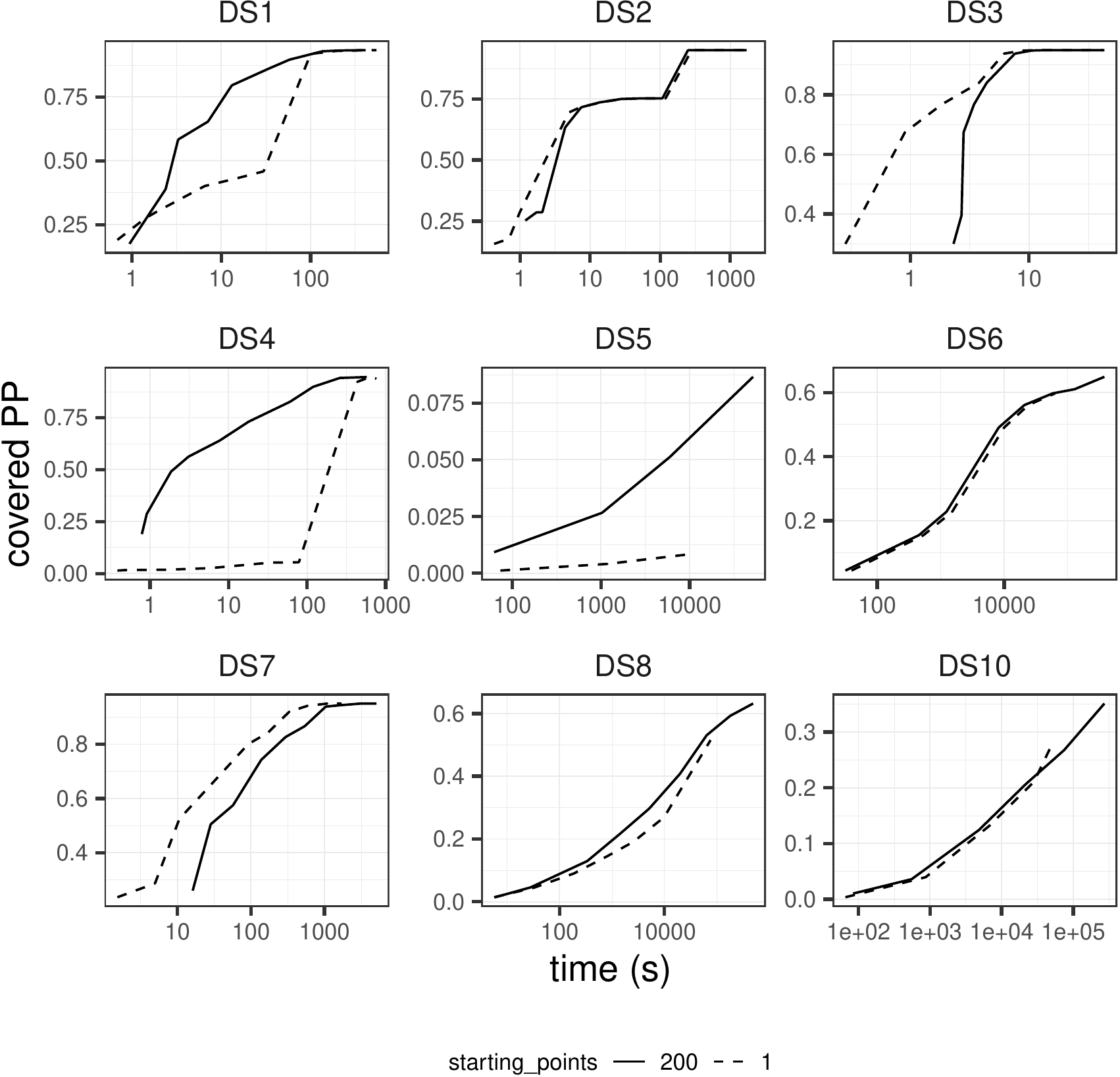}
	\caption{The cumulative posterior probability of HLSs explored by PT with increasing time.}
    \label{fig:coverage_time}
\end{figure}

The time required to explore the HLS to a given threshold varies with the shape of the posterior distribution (Fig.~\ref{fig:coverage_time_subset}; full results in online supplemental Fig.~\ref{fig:coverage_time}).
In general, the time required grew exponentially as the negative threshold decreased.
Using multiple starting points carries only a small performance penalty beyond their initial calculation time.
As such, multiple starting points greatly reduced the time required to cover a given proportion of the credible set on peaky datasets.

\begin{figure}[H]
	\centering
	\includegraphics[width=\textwidth]{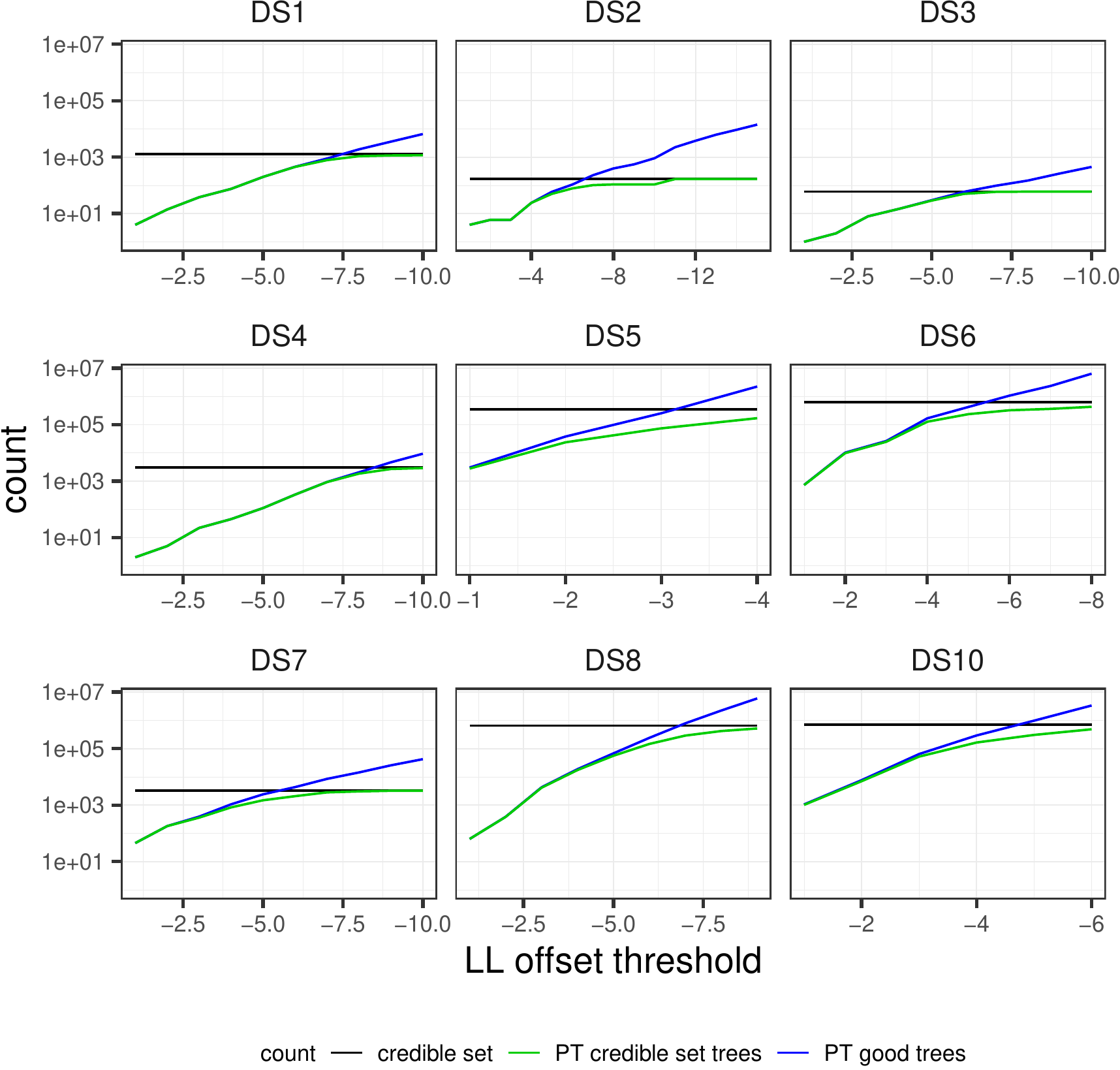}
	\caption{Raw counts of covered and not credible topologies by log-likelihood thresholds.}
	\label{fig:coverage_count}
\end{figure}

The number of visited topologies inside and outside of the credible set both increase with the likelihood threshold (online supplemental Fig.~\ref{fig:coverage_count}).
The majority of HLS topologies found under a small threshold were also in the credible set.
In most of our tests the number of low probability topologies that must be visited to fully explore the 95\% credible set is of the same order of magnitude as the number of visited high probability topologies.
However on datasets with large flat posteriors, such as DS5, the number of low probability topologies was within two orders of magnitude of the number of high probability topologies at the tested thresholds.

\begin{figure}\arxiv{[p]}
	\centering
	\hspace{\stretch{1}}
	\subfigure[\label{fig:time_merge}]{\includegraphics[width=0.37\textwidth]{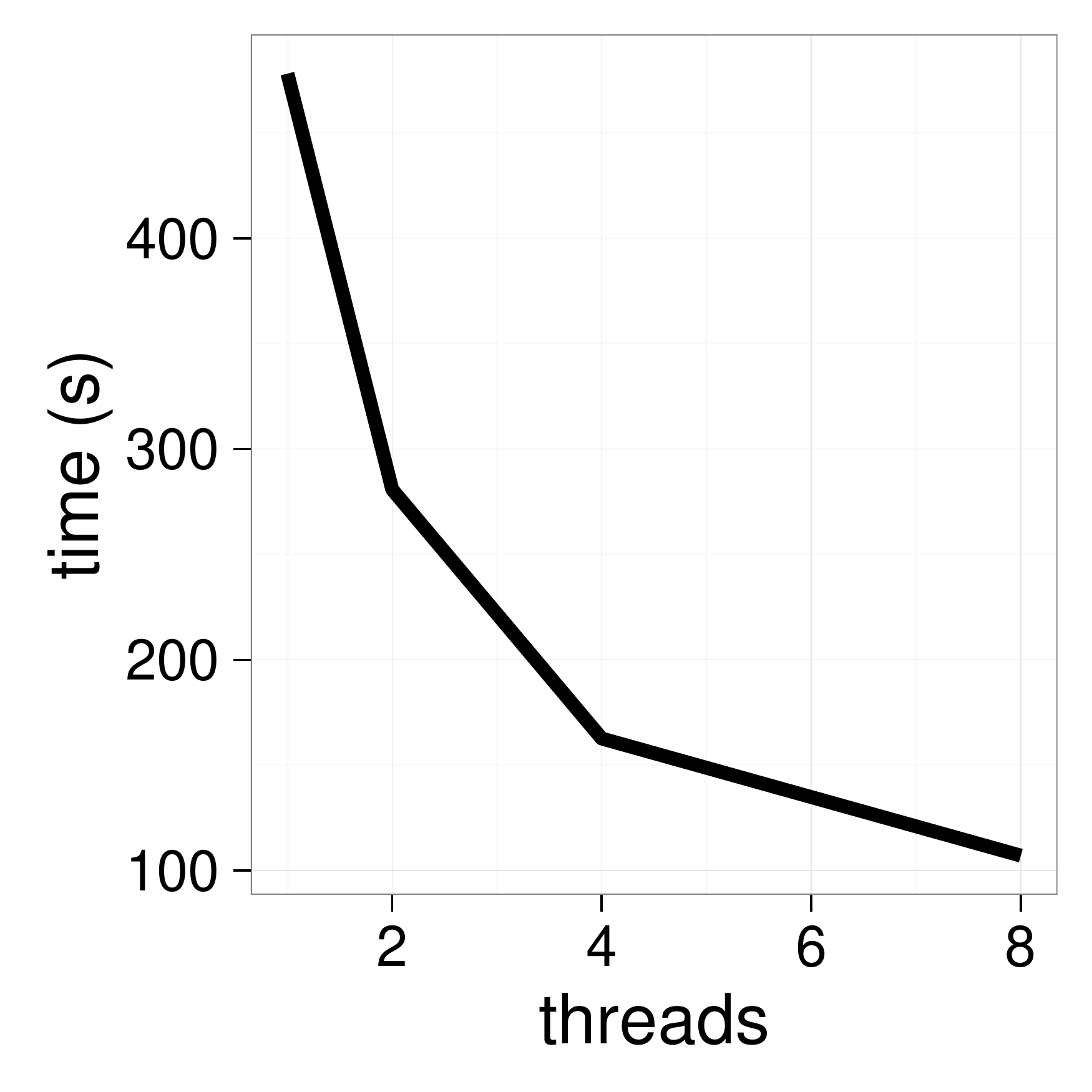}}
	\hspace{\stretch{2}}
		\subfigure[\label{fig:coverage_ds1}]{\includegraphics[width=0.4\textwidth]{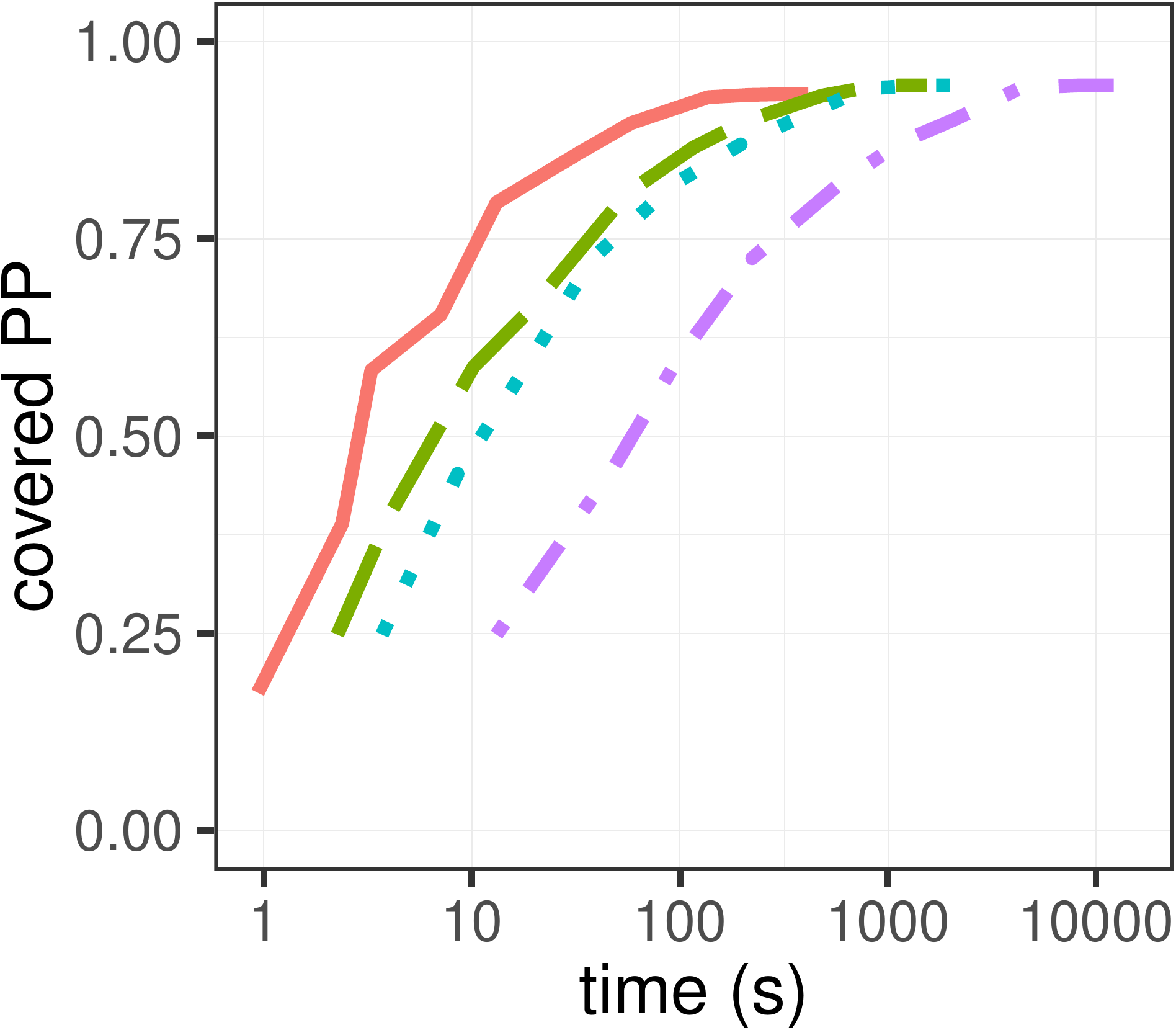}}
	\hspace{\stretch{1}}
	\caption{
    (a) The mean running time improvement from multithreading on DS1 with 200 starting points and a threshold of 10 log-likelihood units.
    (b) The running time improvement from optimizing branch lengths on DS1 with a limited radius of 0 (solid), 1 (dashed), 2 (dotted), and full branch optimization (dotdashed).}
	\label{fig:time_comparison}
\end{figure}

\begin{figure}\arxiv{[p]}
	\centering
	\includegraphics[width=0.8\textwidth]{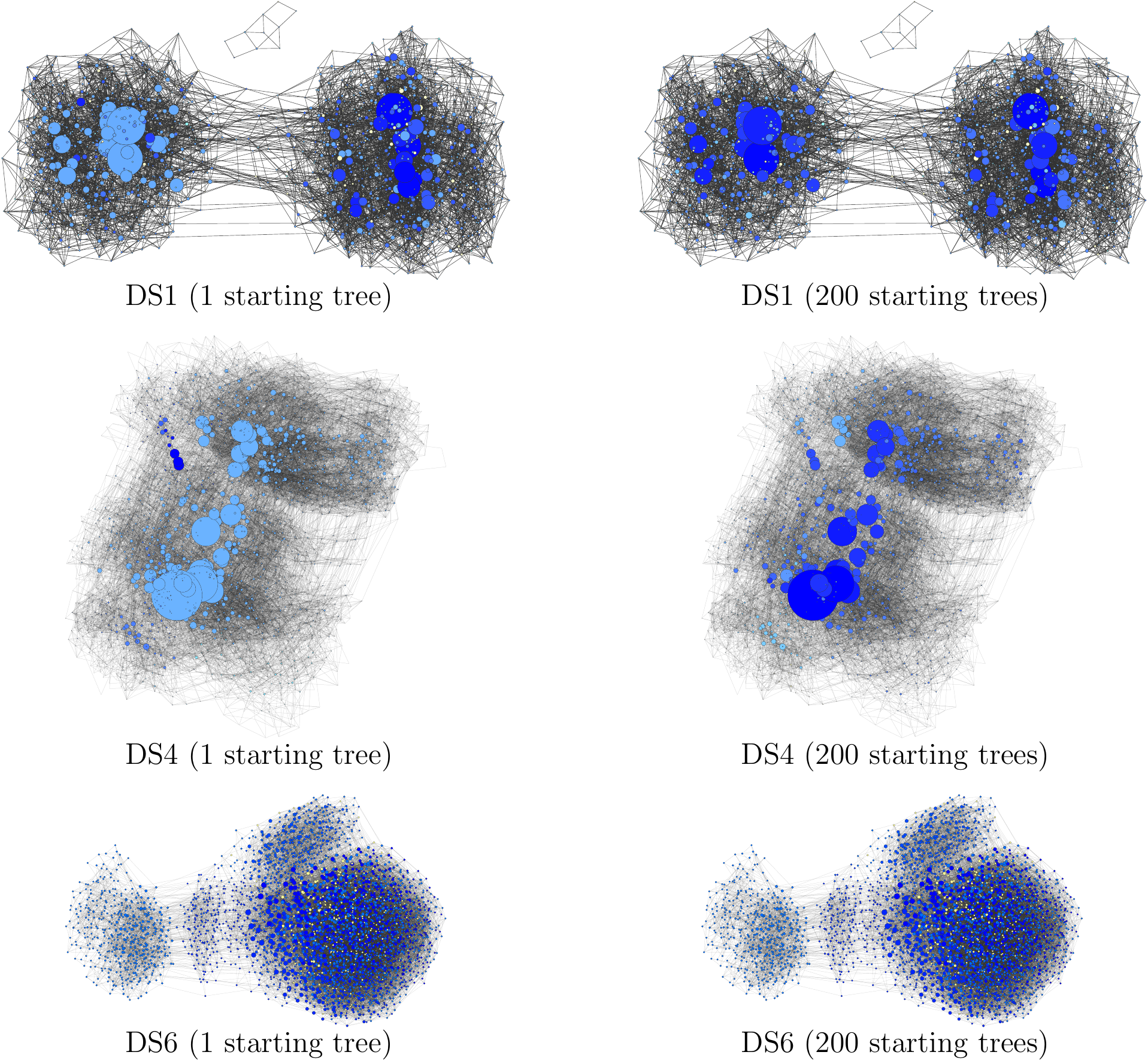}
	\caption{SPR graphs built from MrBayes credible sets of DS1, DS4, and DS6 and colored according to PT exploration.
    Topologies found by PT with successively more negative log-likelihood thresholds are shown on a dark to light blue scale.
		Scales vary per dataset with respect to the most negative computed threshold.
    Yellow means not found.
		}
	\label{fig:pt_graphs_1v200}
\end{figure}
Our implementation of direct topology exploration scales well with multiple threads (Fig.~\ref{fig:time_merge}).
Over 10 replicated tests on DS1 with a threshold of -10, the mean speedup with 8 threads was 4.46x.

Single branch optimization of tested trees outperformed full branch optimization by a large margin (Fig.~\ref{fig:coverage_ds1}).
Profiling showed that the majority of PT computation is optimizing the branch lengths of tree topologies to maximize their likelihood and single branch optimization.
Recall that we fully optimize the branch lengths of visited topologies, so using single branch optimization only runs the risk of missing a topology that belongs in the HLS.
Indeed, full branch optimization recovered a small number of extra topologies at a given threshold.
However, given the extra time required to fully optimize the large number of neighboring tested topologies with small likelihoods, it was more beneficial to simply decrease the threshold rather than optimize the length of every tree branch or even every tree branch at a distance of 1 or 2 from the newly introduced branch.

\begin{figure}[H]
	\centering
	\includegraphics[width=\textwidth]{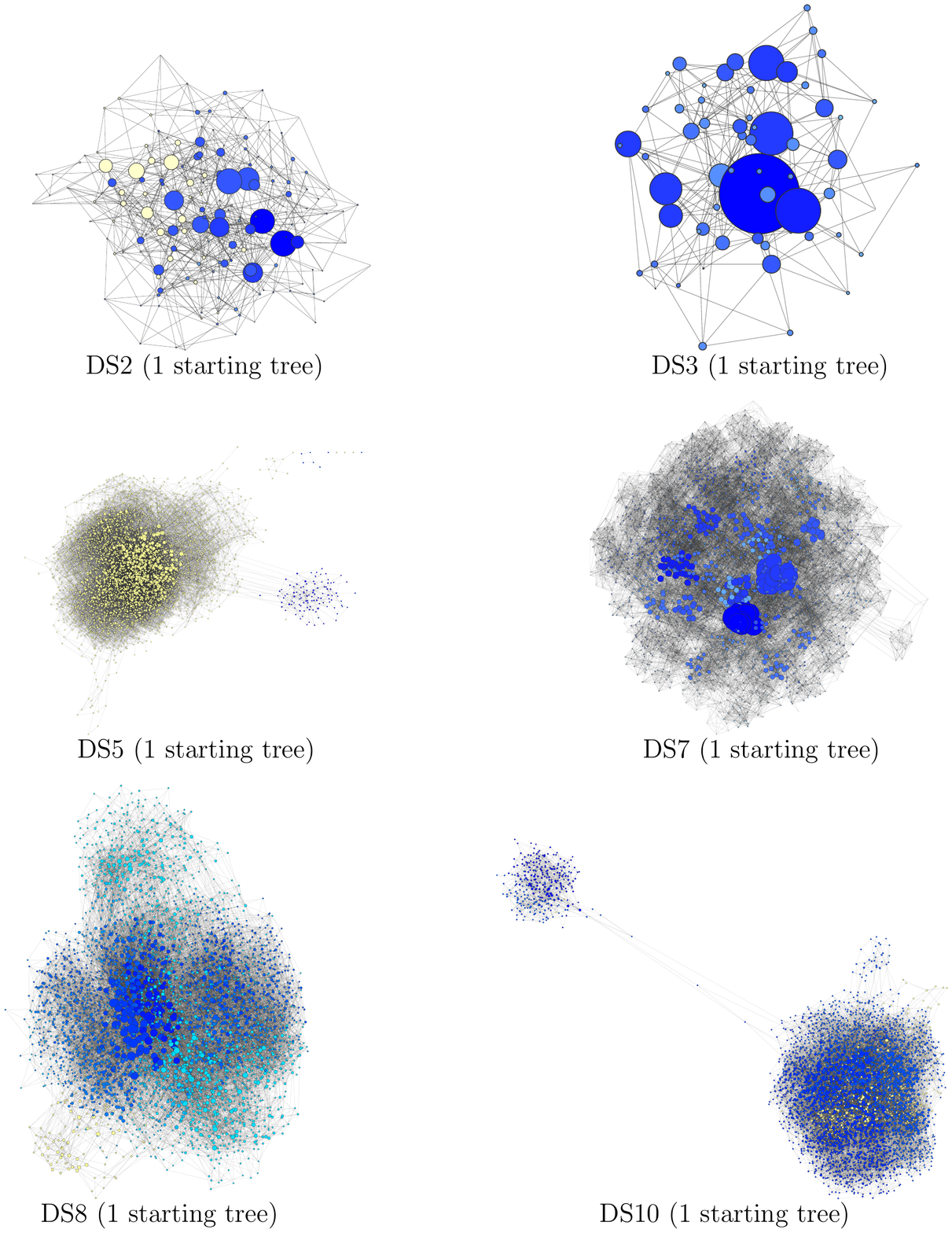}
	\caption{
    SPR graphs of the credible sets of the remaining datasets with 1 starting point.
    Topologies found by PT with successively more negative log-likelihood thresholds are shown on a dark to light blue scale.
    Yellow means not found.
		}
	\label{fig:pt_graphs_single}
\end{figure}

SPR graphs showed that a single starting topology often began exploration from a smaller mode in peaky datasets (Fig~\ref{fig:pt_graphs_1v200}, online supplemental Fig.~\ref{fig:pt_graphs_single}).
These graphs of the MrBayes credible sets with overlayed PT HLS demonstrate the expansion of the HLS at different likelihood thresholds.
In DS1, a search threshold of -7 log-likelihood units is required to bridge the gap between the starting topology and the second peak.
In DS4, the RAxML starting tree was stuck in a poor local optima and a search threshold of -8 log-likelihood units was required to find the majority of the high likelihood topologies.
In both cases, using multiple starting points greatly improved exploration of the HLS at small thresholds: PT runs with 200 starting points have darker blue, showing in more detail how multiple starting points get more good topologies with a higher likelihood threshold.
Surprisingly, none of our 200 starting points landed in the smaller mode of DS6.

\begin{figure}
	\centering
	\includegraphics[width=\textwidth]{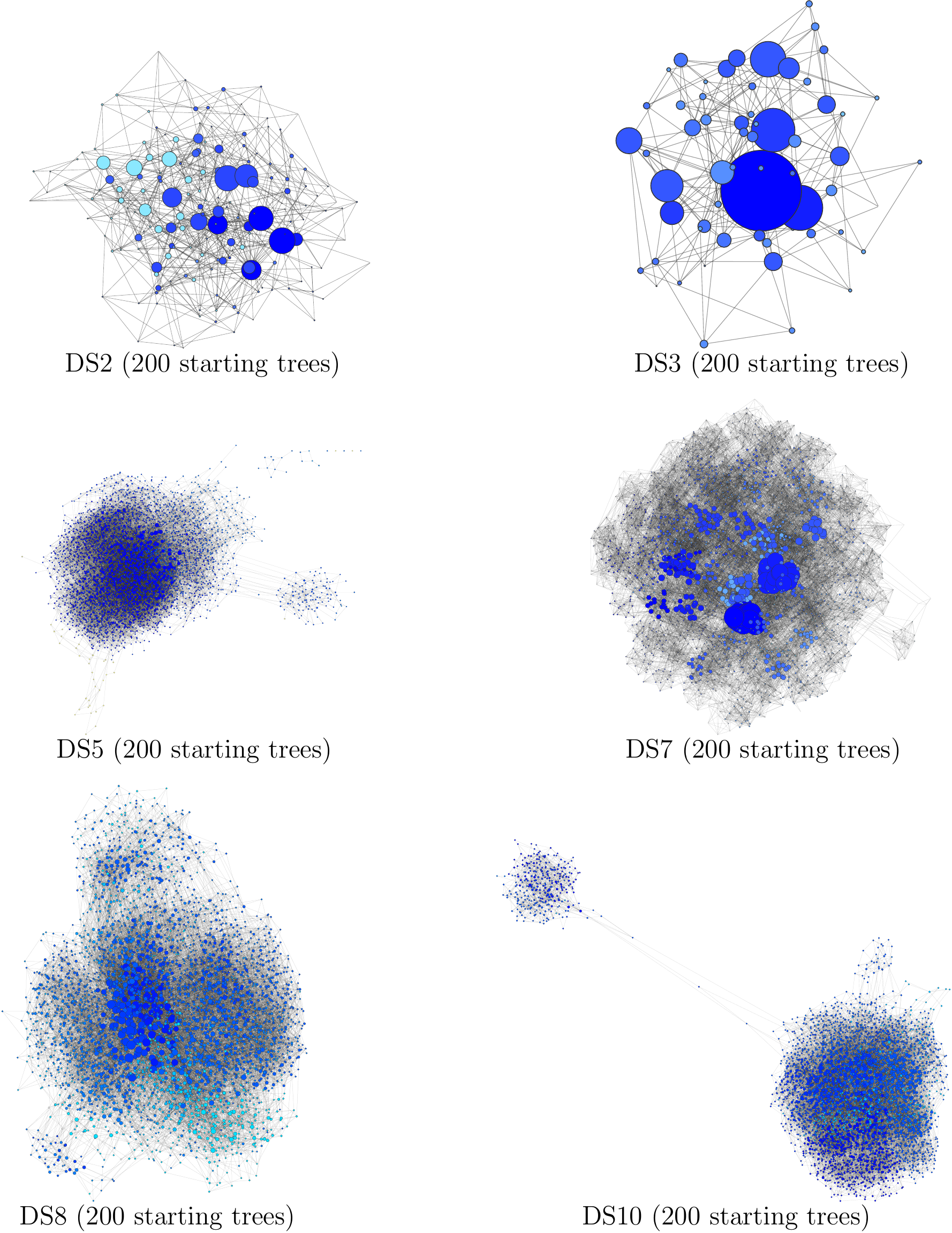}
	\caption{
    SPR graphs of the credible sets of the remaining datasets with 200 starting points.
    Topologies found by PT with successively more negative log-likelihood threshold are shown on a dark to light blue scale.
    Yellow means not found.
    }
	\label{fig:pt_graphs_200}
\end{figure}

SPR graphs of the other six datasets showed a clear expansion of the HLS outward from the most likely topologies (Fig.~\ref{fig:pt_graphs_200}).
As expected from Figure~\ref{fig:coverage} the expansion was similar with 1 or 200 starting points.

\begin{figure}
	\centering
	\includegraphics[width=0.8\textwidth]{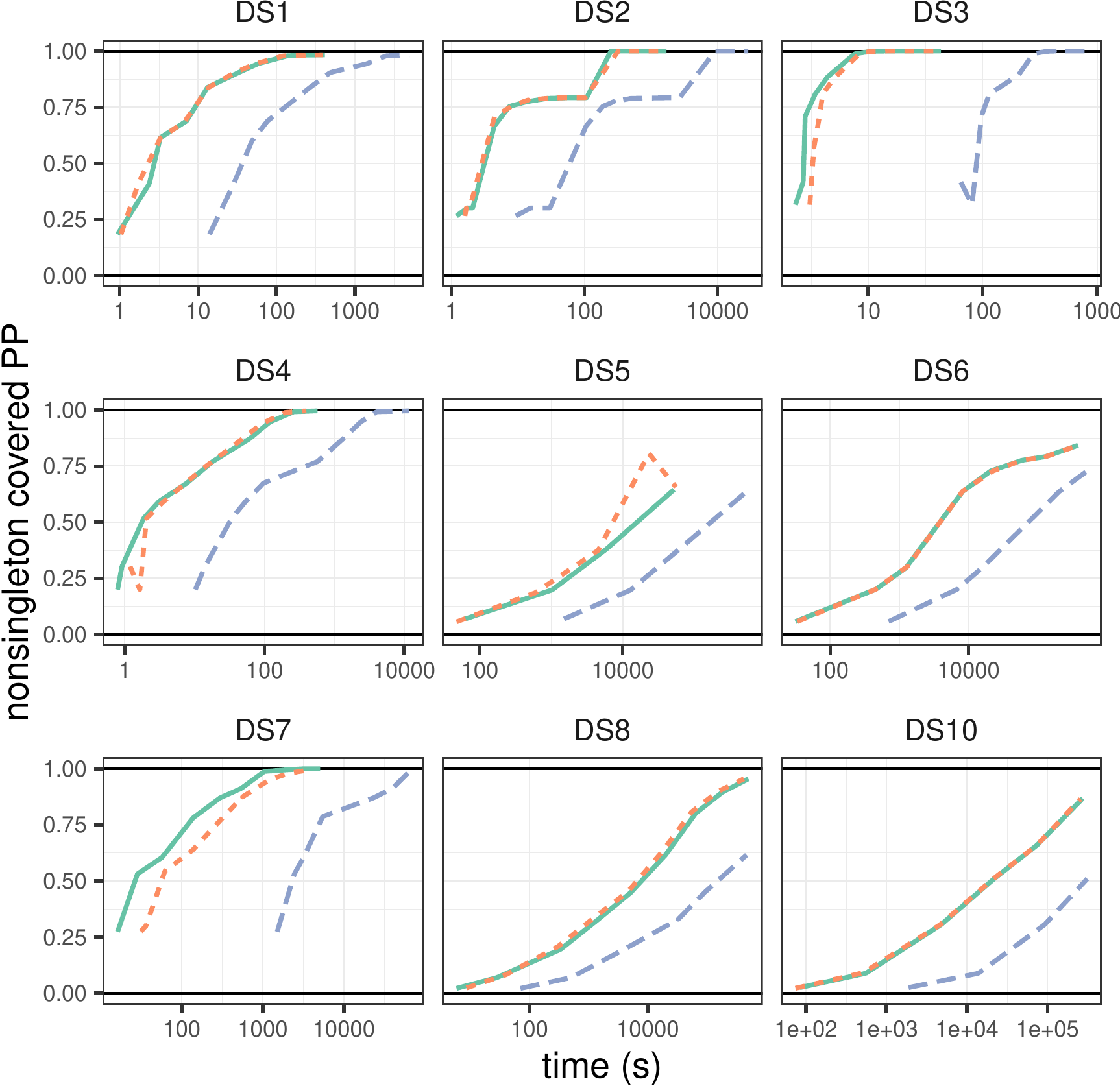}
	\caption{
    The nonsingleton covered PP (from MrBayes) with increasing time.
		Default (solid), MAP (dotted), and model testing (dashed) runs were evaluated with 200 starting points.}
	\label{fig:coverage_normalized}
\end{figure}

PT quickly found the majority of the nonsingleton posterior probability (Fig.~\ref{fig:coverage_normalized}).
Restricting our comparisons to topologies in the credible set which were sampled at least twice showed that PT found all or most such topologies, even on difficult flat datasets like DS5.
We were surprised to find that only 13.6\% of our computed ``95\% credible set'' for DS5 was composed of topologies sampled more than once in our 10 different golden run replicates.
On such datasets the HLS may be a small fraction of the MrBayes credible set, complicating the comparison of HLS and credible sets.
This suggests that, even with golden runs, the posterior probabilities of singleton topologies are estimated poorly.

Using MAP in place of maximized likelihood or optimizing the gamma parameter of the JC+$\Gamma$ model for each visited topology had little effect on which topologies were found for a given threshold (Fig.~\ref{fig:coverage_normalized}).
Model optimization, in particular, required a large amount of extra computation that would be better spent exploring to a greater depth with a larger threshold.

\subsection*{Normalized Likelihoods Approximate Topology Posteriors}

\begin{figure}
	\centering
	\includegraphics[width=\textwidth]{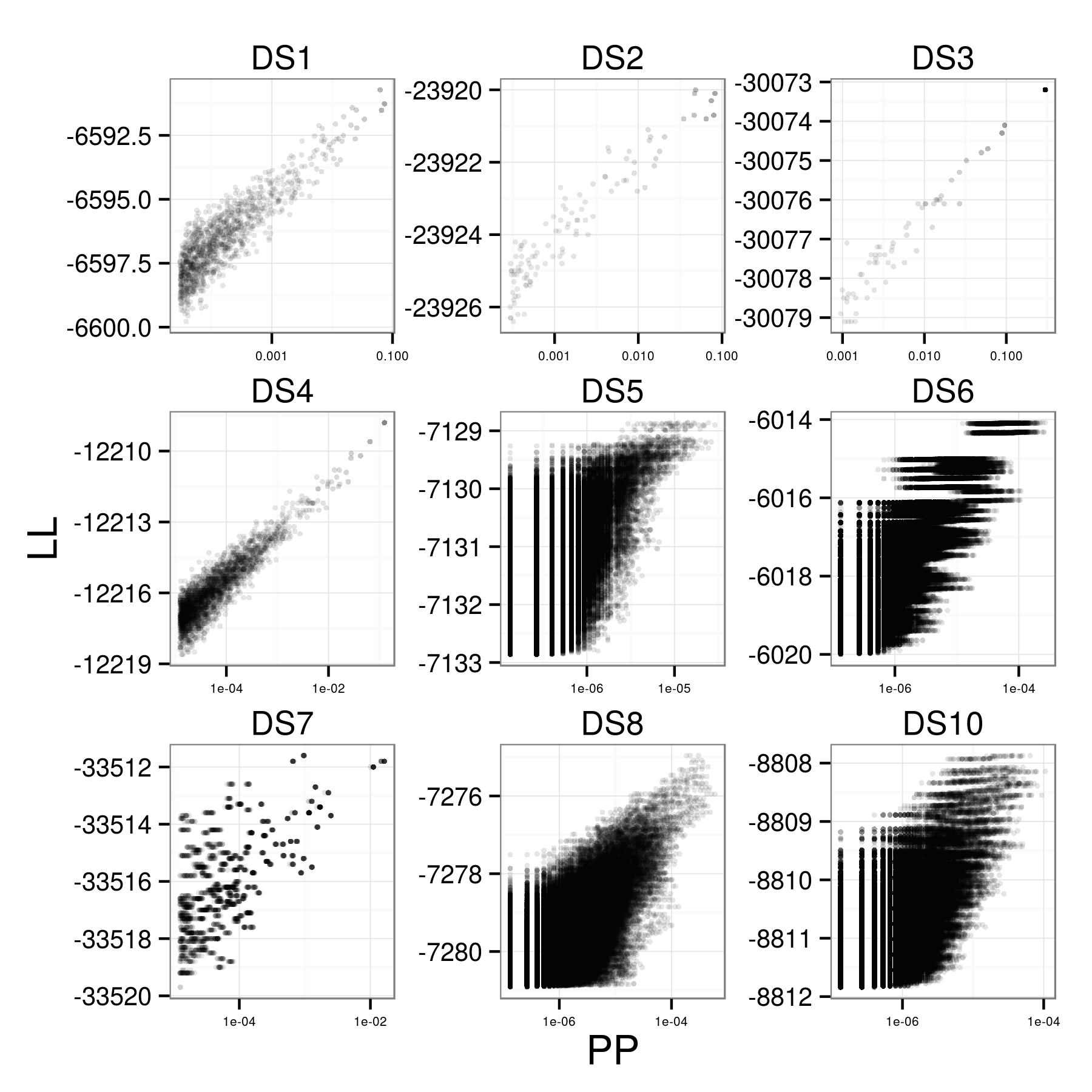}
	\caption{
    Individual topology LL vs MrBayes PP for credible set topologies using 200 starting points.
    Runs are taken at the most negative log-likelihood threshold completed for standard, MAP, and model testing runs.}
	\label{fig:pp_ll_merged_default_logx}
\end{figure}

The probability-normalized likelihoods (i.e.\ likelihoods divided by their sum) of branch-length-optimized tree likelihoods were a surprisingly good indicator of their posterior probability (Fig.~\ref{fig:pp_ll_merged_default_logx}).
This relation was strongest on datasets with a small credible set.
Moreover, there was a close correspondence between the sum of the maximized likelihoods and the sum of the posterior probabilities of an HLS.

\begin{figure}[H]
	\centering
	\includegraphics[width=\textwidth]{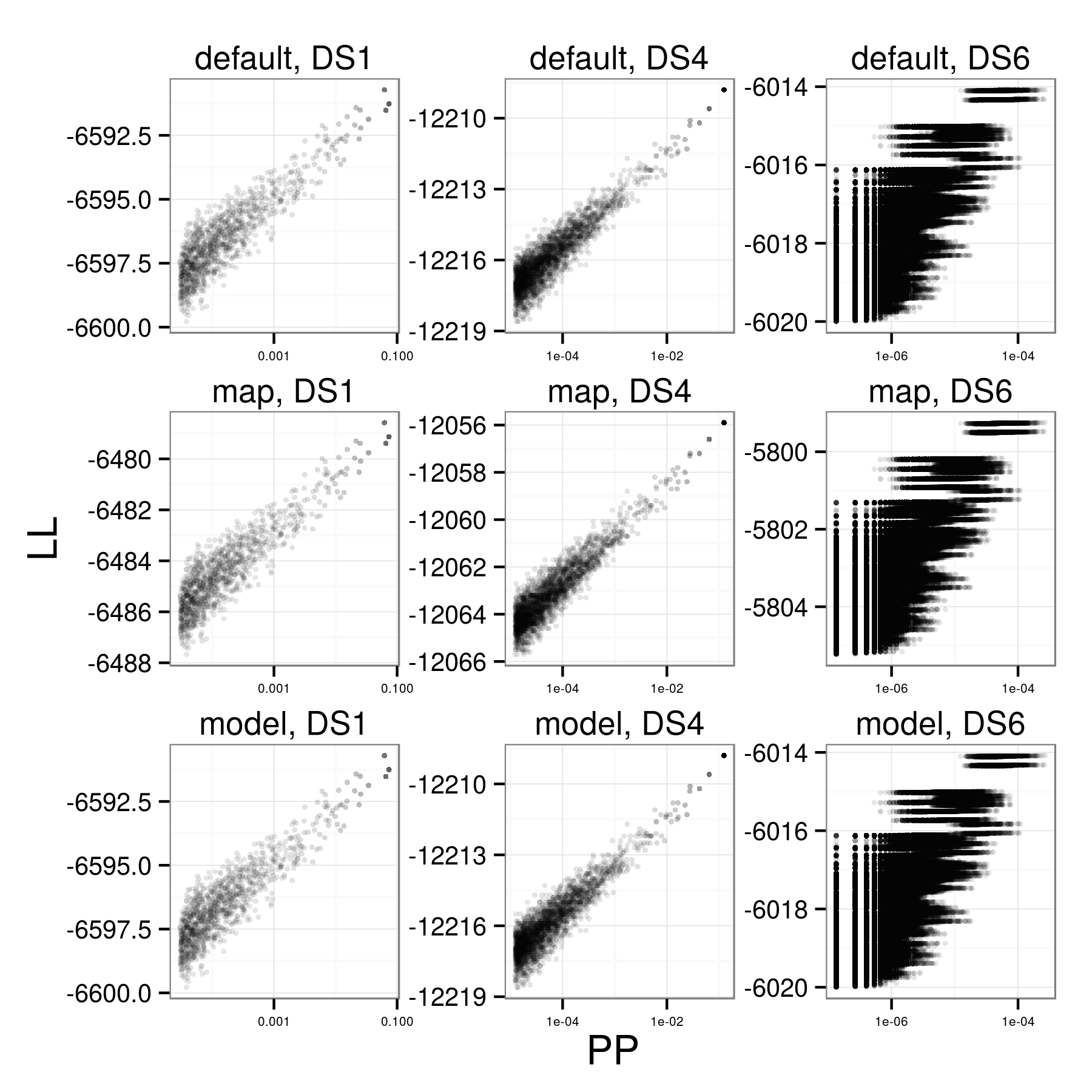}
	\caption{Individual tree topology LL vs PP for credible set topologies with default, MAP and model testing on DS1, DS4 and DS6 with 200 starting points.}
	\label{fig:pp_ll_merged_logx}
\end{figure}

Our three different methods of estimating topology likelihoods showed similar patterns relative to posterior probabilities (online supplemental Fig.~\ref{fig:pp_ll_merged_logx}).
Using a fixed gamma parameter with standard maximum likelihood, MAP, or optimizing the gamma parameter with maximum likelihood made little difference in the datasets we studied.

\begin{figure}
	\centering
	\subfigure[DS1\label{fig:split_rmsd_ds1}]{\includegraphics[width=0.3\textwidth]{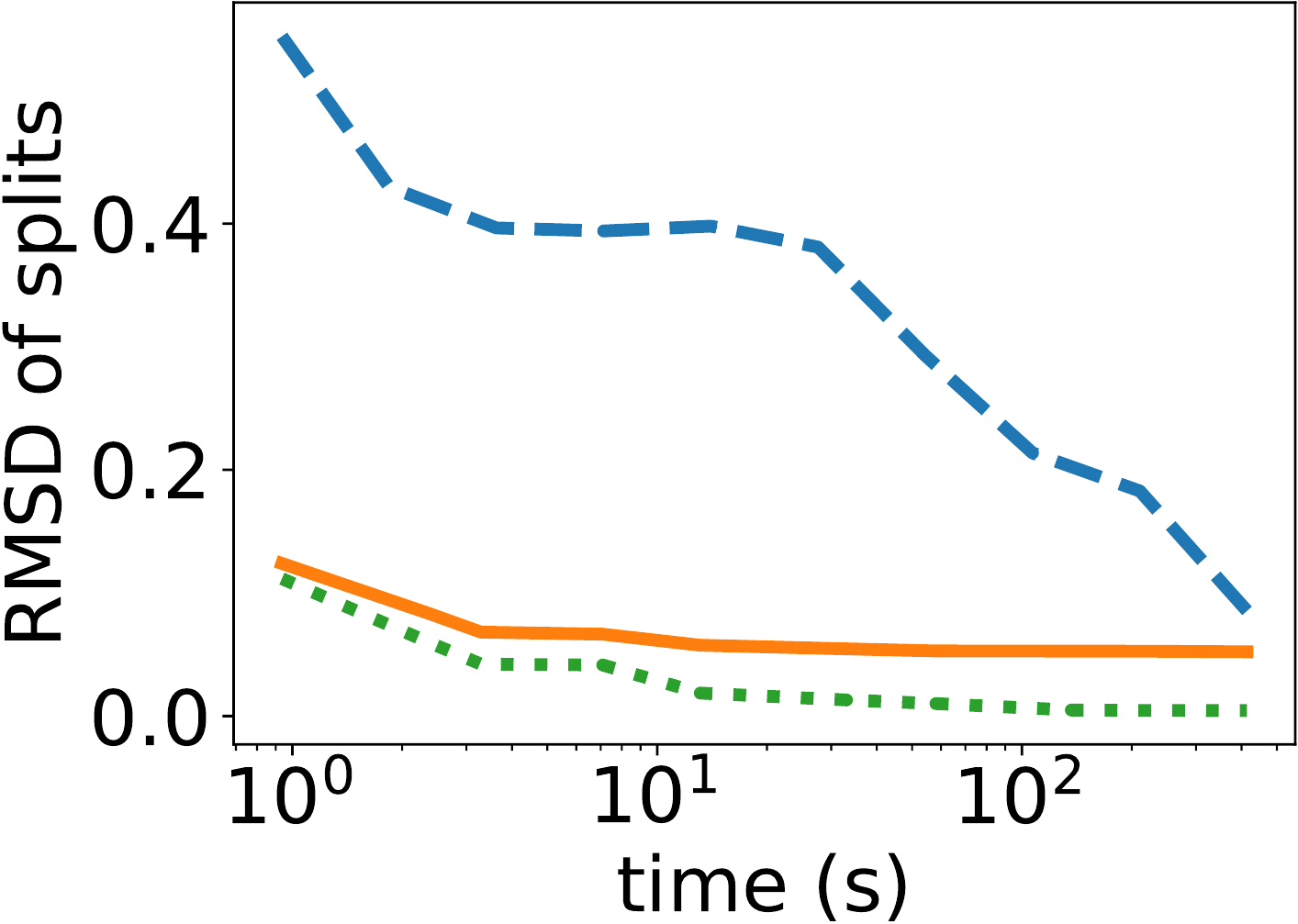}}
	\subfigure[DS4\label{fig:split_rmsd_ds4}]{\includegraphics[width=0.3\textwidth]{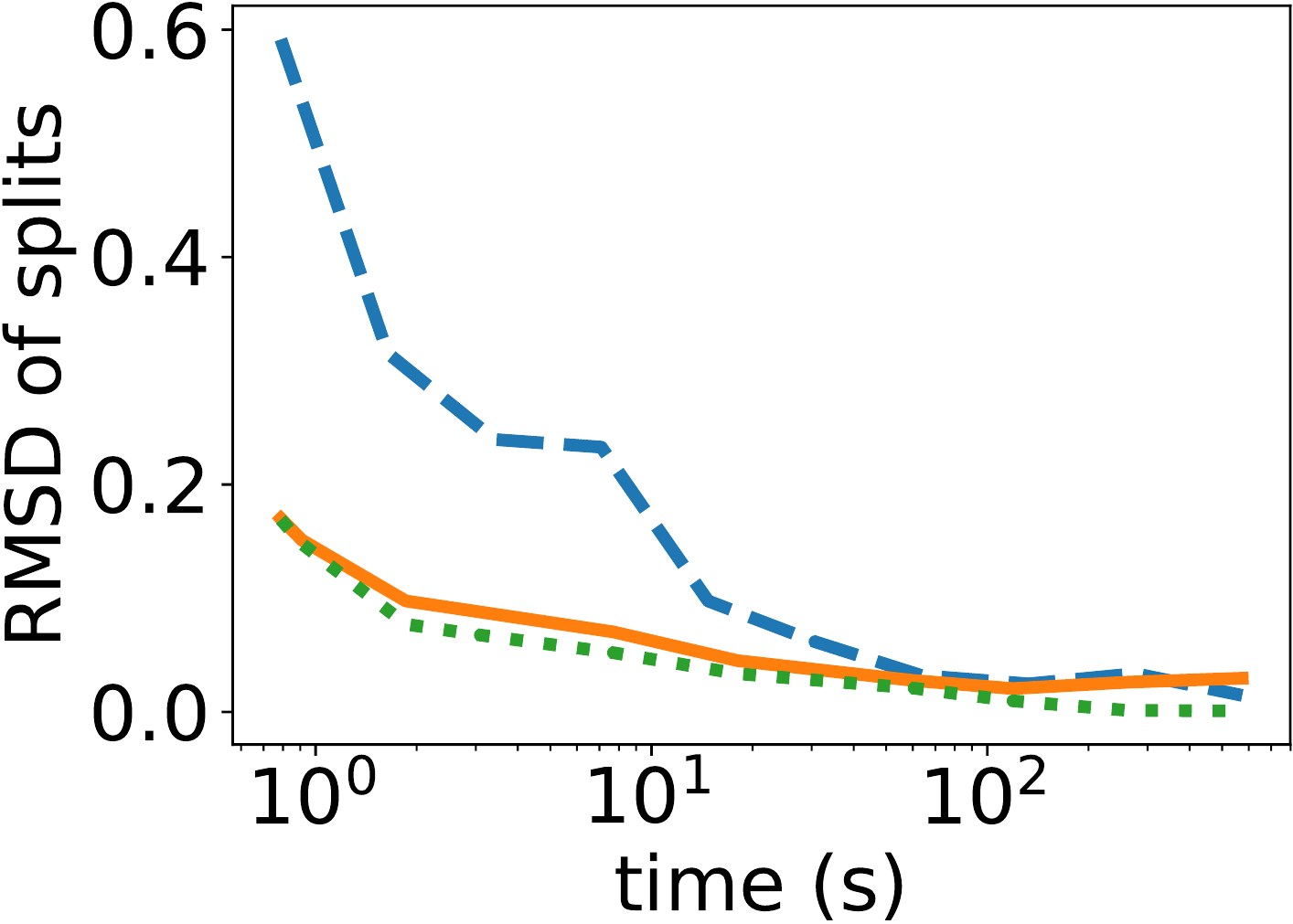}}
	\subfigure[DS6\label{fig:split_rmsd_ds6}]{\includegraphics[width=0.3\textwidth]{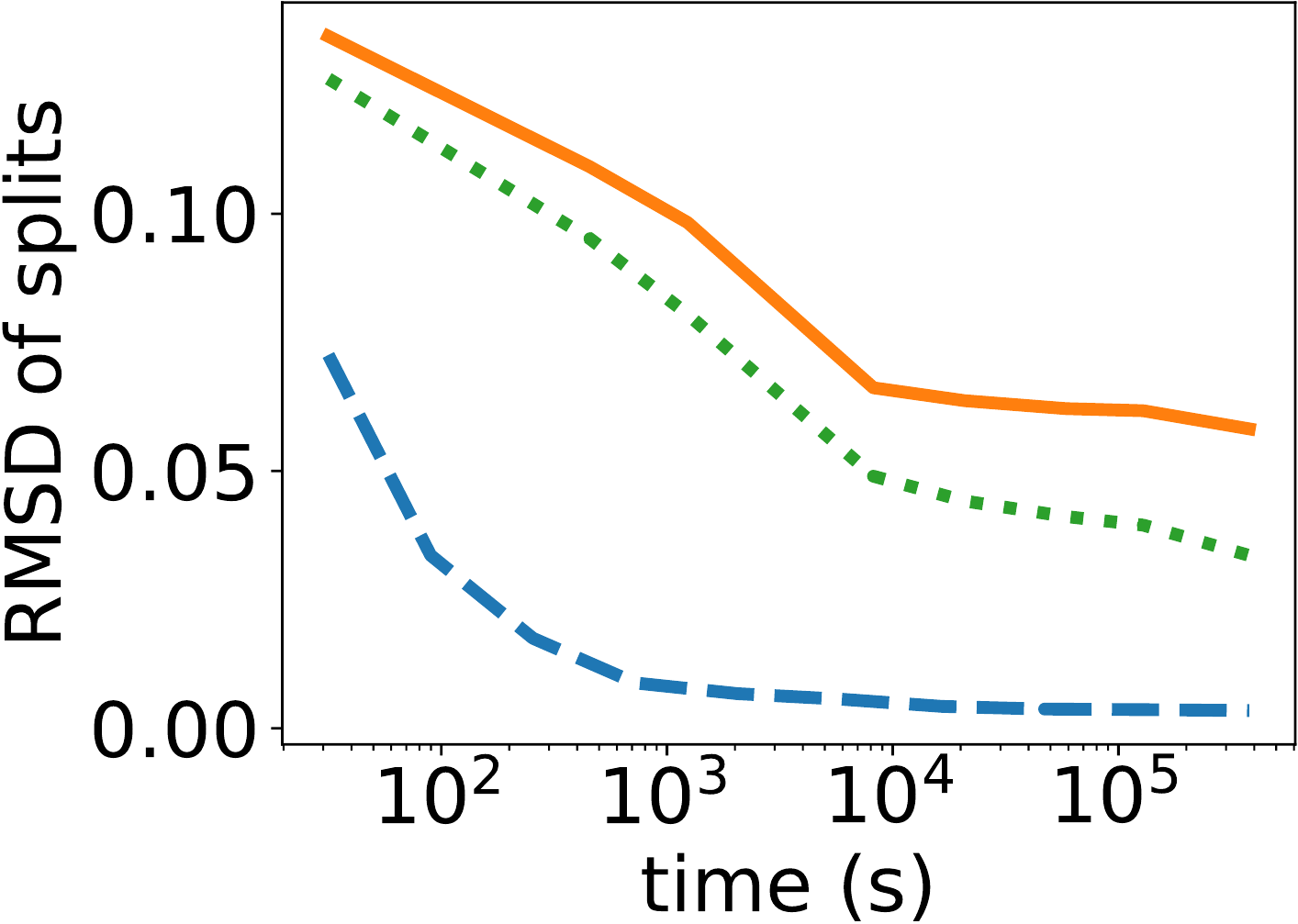}}
	\caption{Comparison of RMSD of split frequencies from aggregated golden runs over time for MrBayes (dashed), PT (solid), and PT assuming MrBayes posterior probabilities (dotted). PT results used 200 starting points.}
	\label{fig:split_rmsd}
\end{figure}

\begin{figure}[H]
	\centering
	\subfigure[DS2\label{fig:split_rmsd_ds2}]{\includegraphics[width=0.3\textwidth]{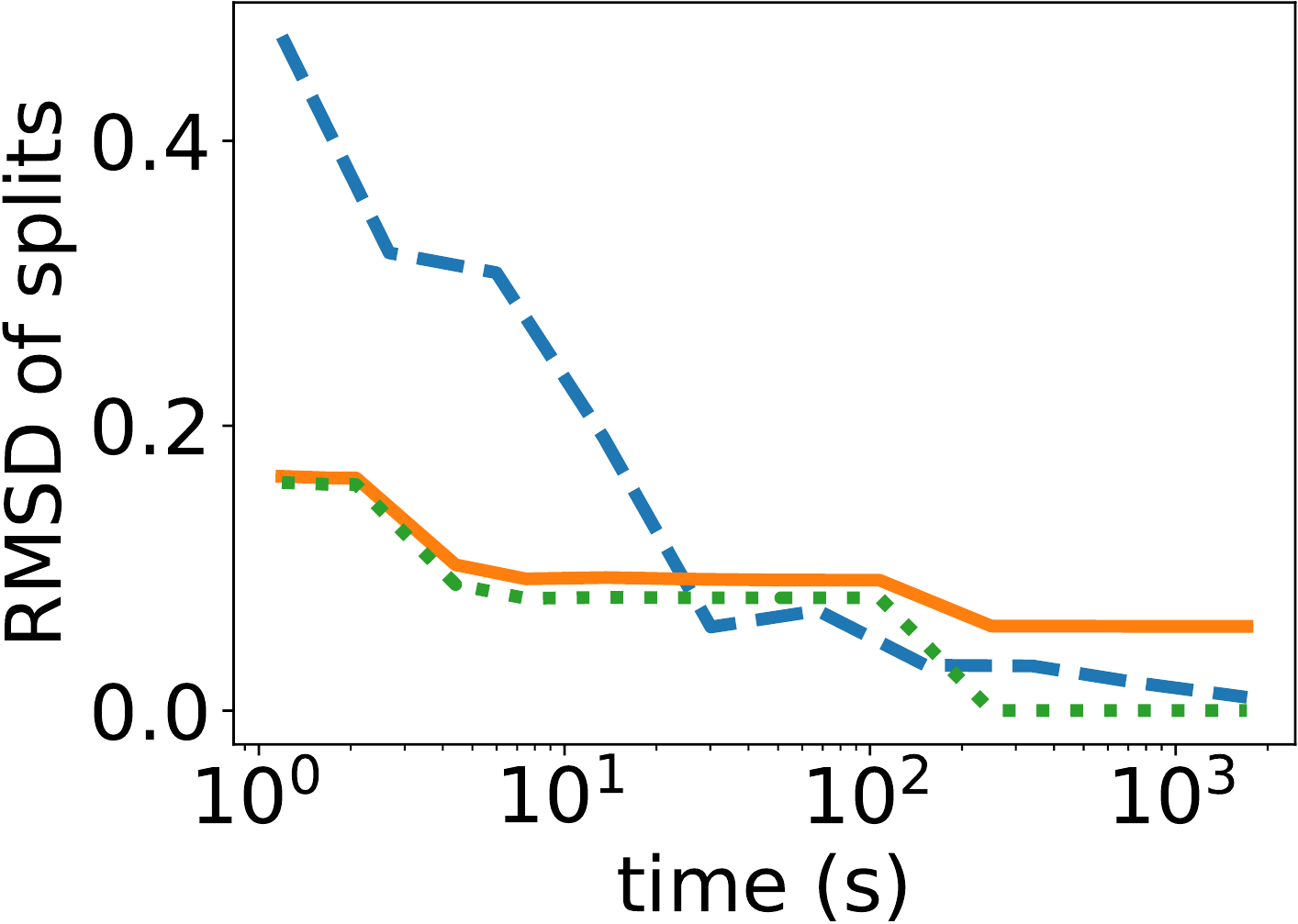}}
	\subfigure[DS3\label{fig:split_rmsd_ds3}]{\includegraphics[width=0.3\textwidth]{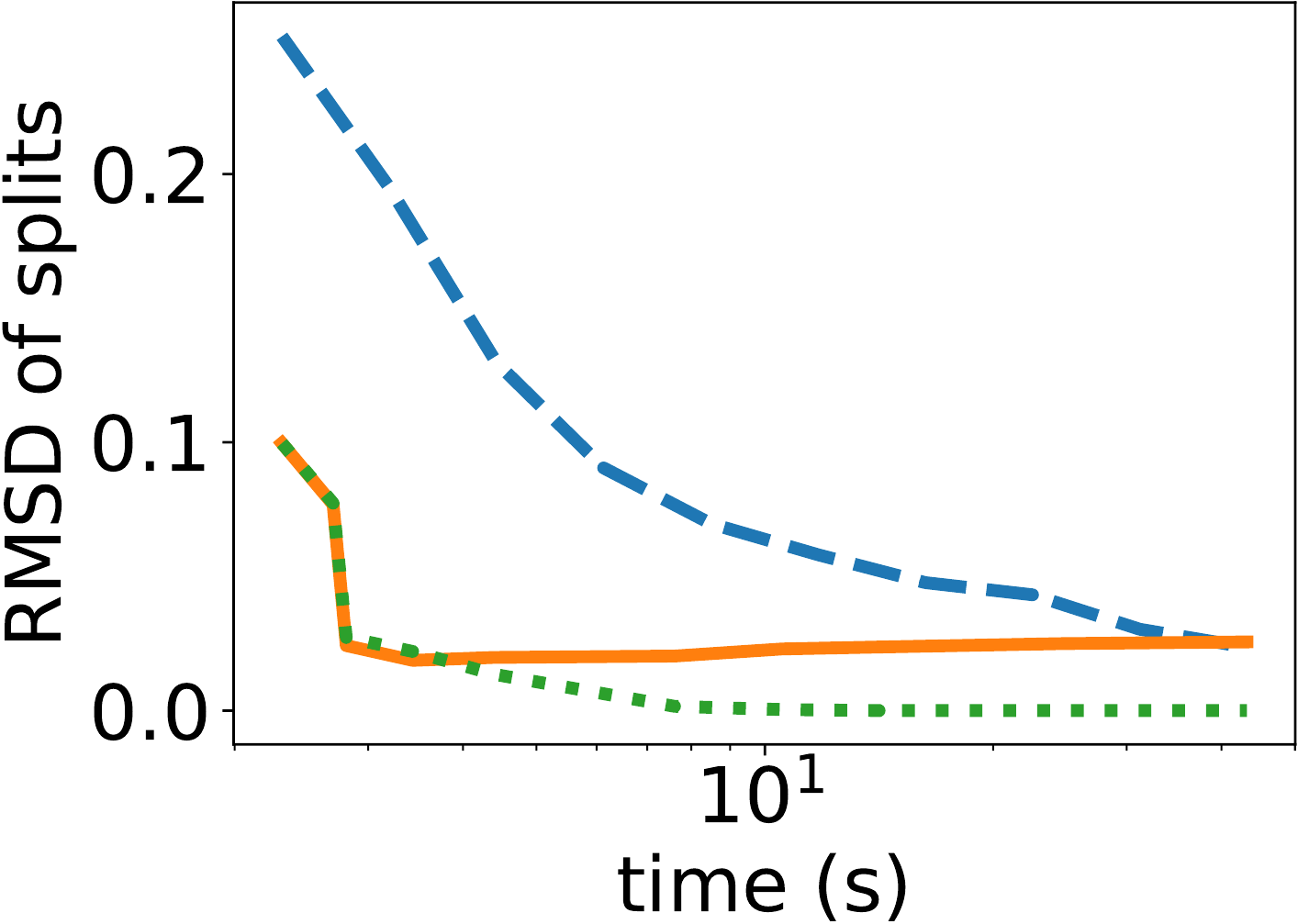}}
	\subfigure[DS5\label{fig:split_rmsd_ds5}]{\includegraphics[width=0.3\textwidth]{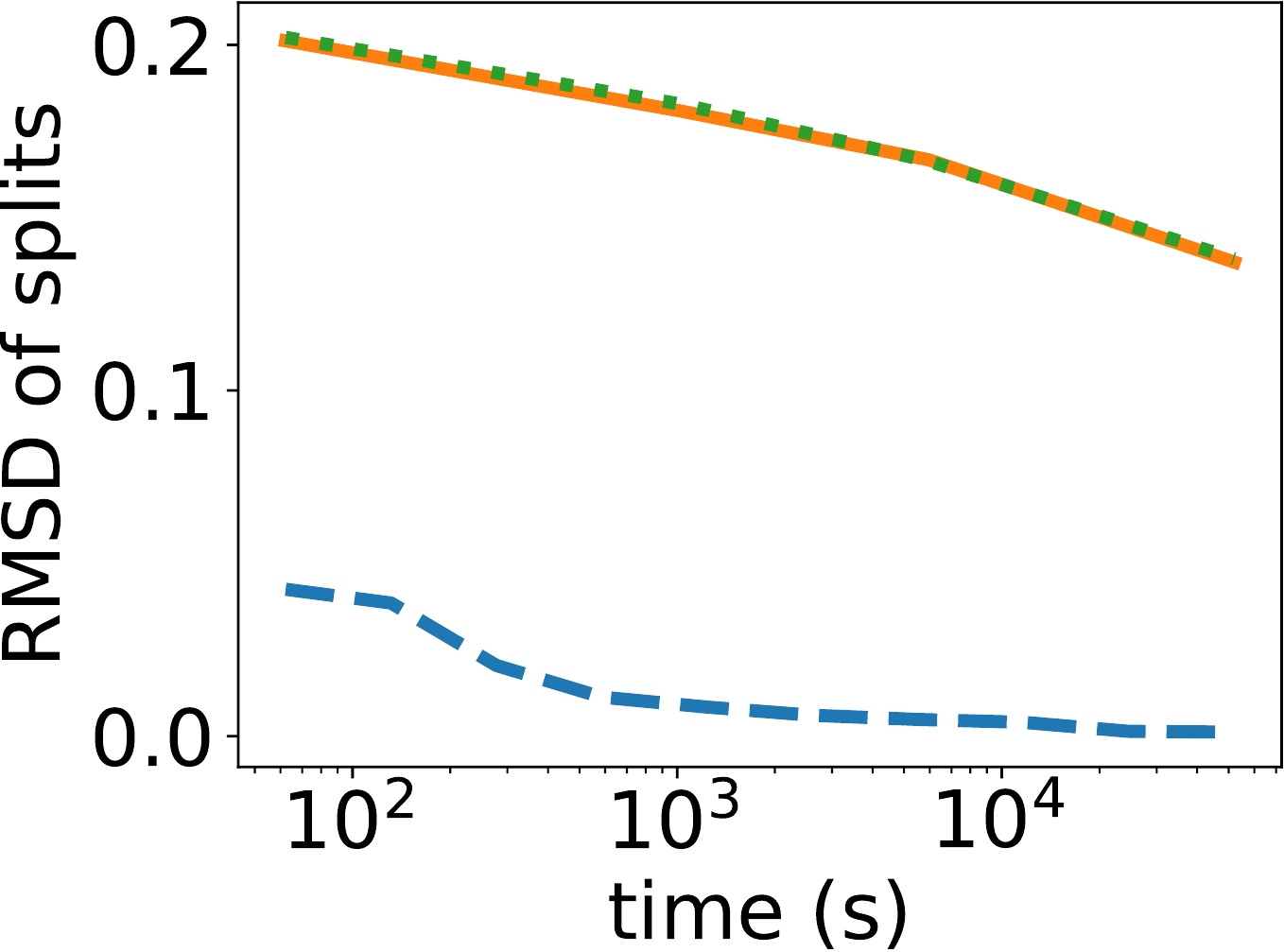}}
	\subfigure[DS7\label{fig:split_rmsd_ds7}]{\includegraphics[width=0.3\textwidth]{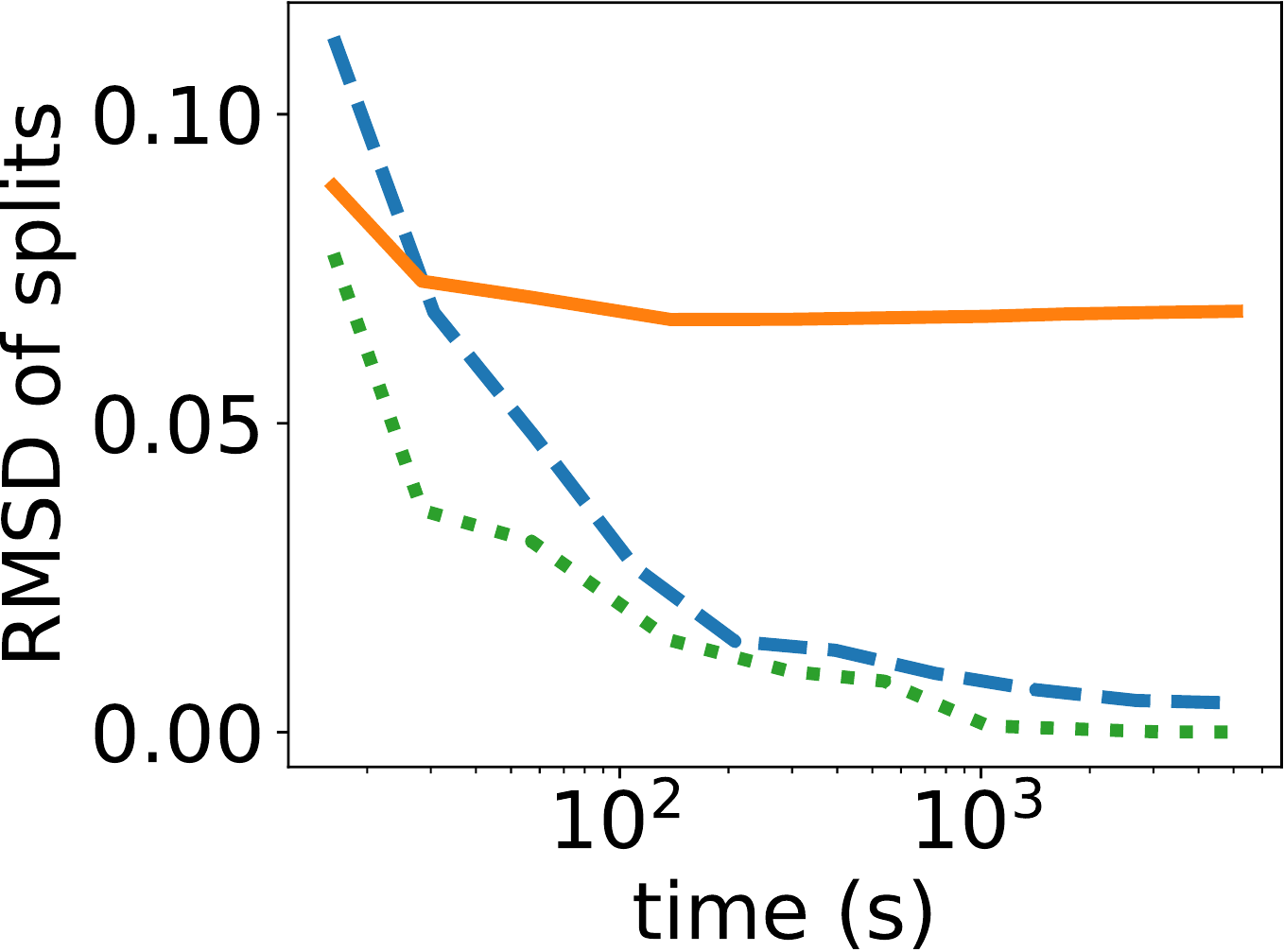}}
	\subfigure[DS8\label{fig:split_rmsd_ds8}]{\includegraphics[width=0.3\textwidth]{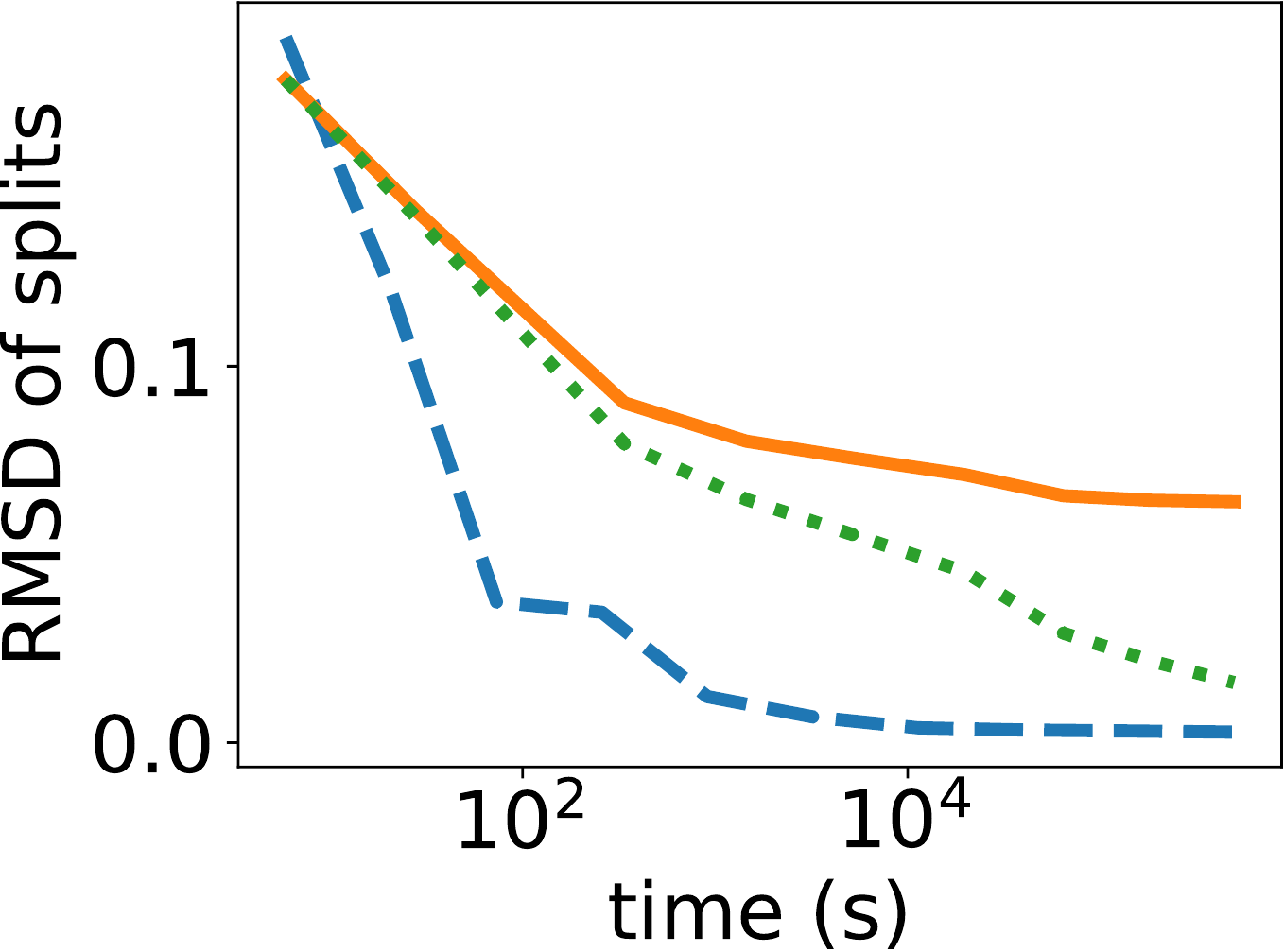}}
	\subfigure[DS10\label{fig:split_rmsd_ds10}]{\includegraphics[width=0.3\textwidth]{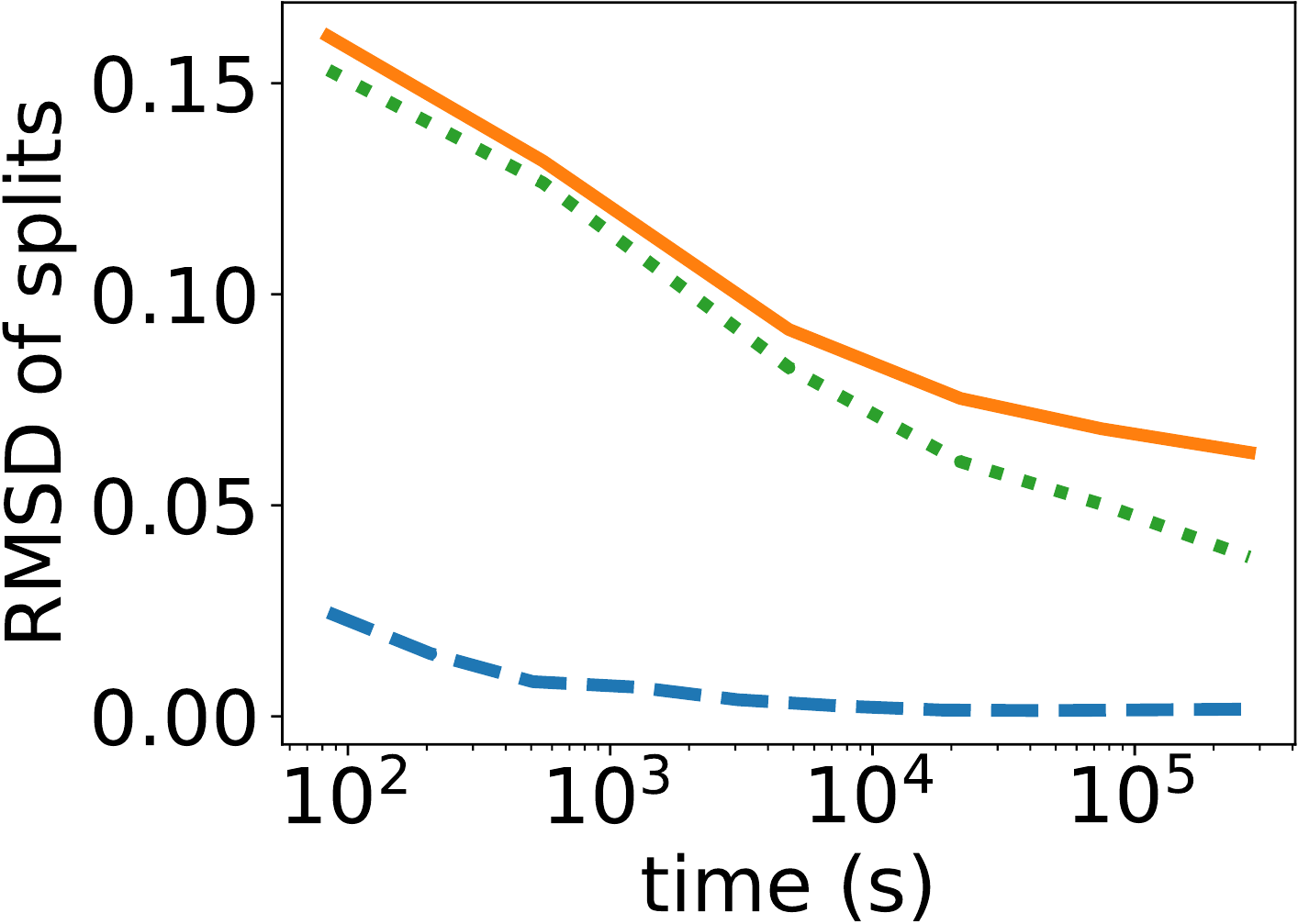}}
	\caption{Comparison of RMSD of split frequencies from aggregated golden runs over time for MrBayes (dashed), PT (solid), and PT assuming MrBayes posterior probabilities (dotted).}
	\label{fig:split_rmsd_supp}
\end{figure}

Split frequencies estimated using maximized likelihood to approximate the marginal likelihood of the HLS rapidly approached the golden run split frequencies (Fig.~\ref{fig:split_rmsd}, online supplemental Fig.~\ref{fig:split_rmsd_supp}).
Here we compare the RMSD between PT split frequencies and the golden run split frequencies over time, as well as the RMSD between single run MrBayes split frequencies and the golden run split frequencies over time.
On datasets with a small credible set the RMSD of PT split frequencies reached our goal of 0.05 more quickly than the RMSD of single run MrBayes split frequencies.
On datasets with larger credible sets PT required more time than MrBayes and approached the golden run split frequencies but did not meet our goal of 0.05.
PT would require more time or memory to fully explore these datasets using maximum likelihood for the topology posterior probability estimate.
We also wished to understand if the deviation to ground truth was because PT exploration strategy was missing trees, or if our use of maximized likelihood to approximate posterior probability was too coarse.
For this reason we also calculated split frequencies using the golden run posterior probabilities attached to the trees in the HLS.
Computing split frequencies in this way was sufficient to reach an RMSD of 0.05 in every dataset except DS5 (dotted lines, Fig.~\ref{fig:split_rmsd}), emphasizing the need for good marginal likelihood estimation procedures beyond maximized likelihood.

\begin{figure}
	\centering
	\subfigure[DS1\label{fig:split_scatter_ds1}]{\includegraphics[width=0.3\textwidth]{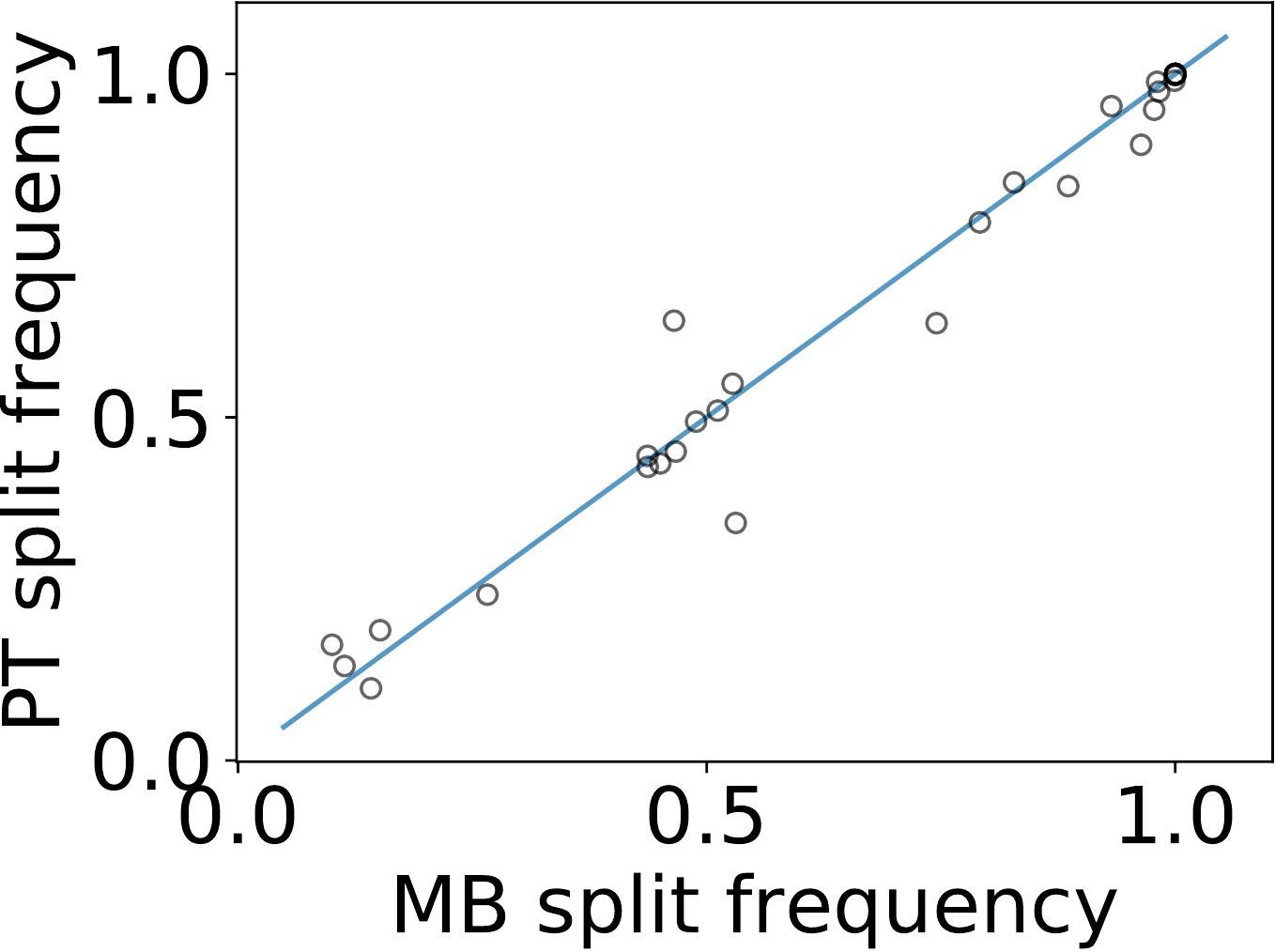}}
	\subfigure[DS4\label{fig:split_scatter_ds4}]{\includegraphics[width=0.3\textwidth]{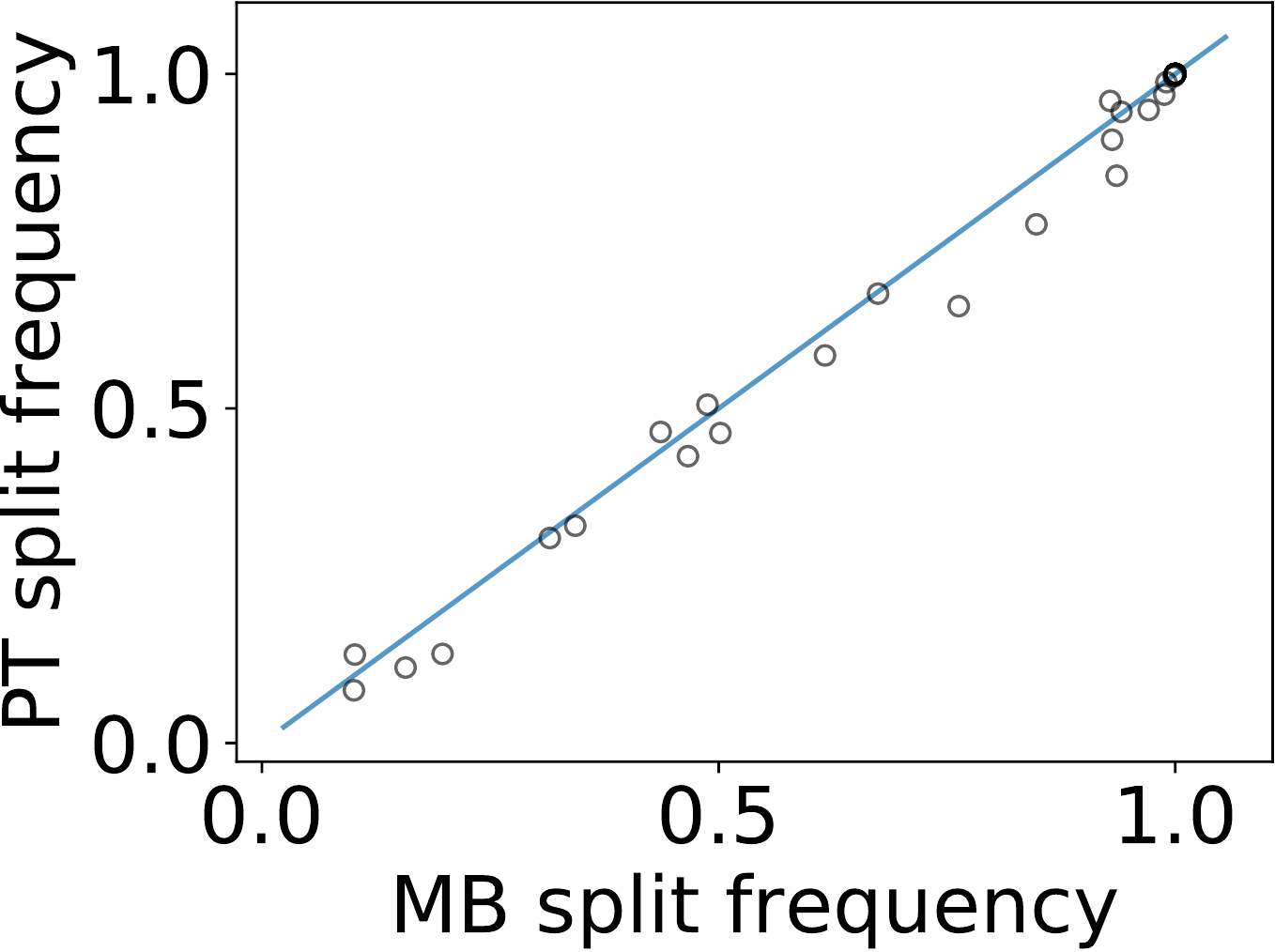}}
	\subfigure[DS6\label{fig:split_scatter_ds6}]{\includegraphics[width=0.3\textwidth]{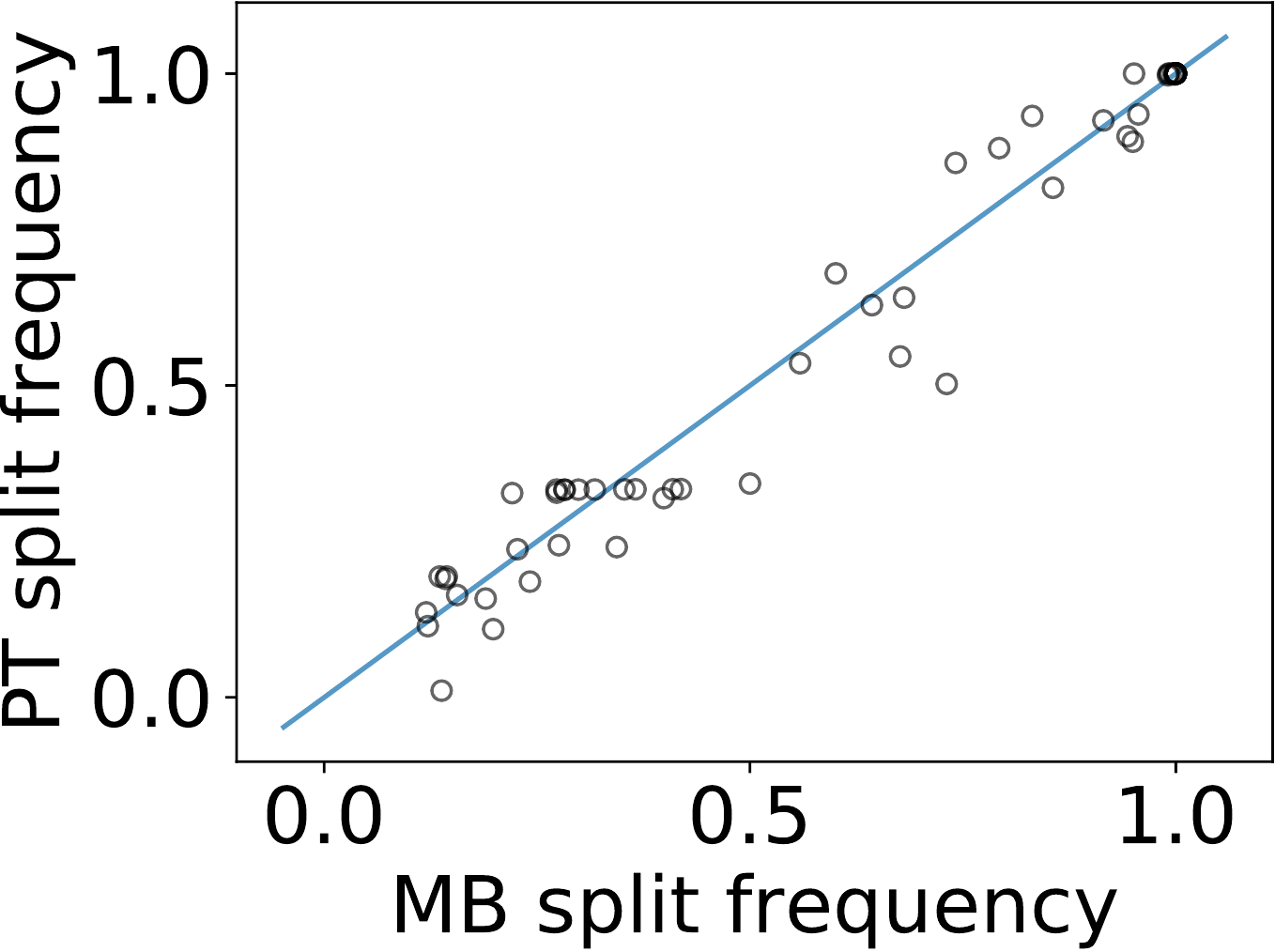}}
	\caption{Comparison of split frequencies from aggregated golden runs over time for MrBayes and PT.}
	\label{fig:split_scatter}

\end{figure}

\begin{figure}[H]
	\centering
	\subfigure[DS2\label{fig:split_scatter_ds2}]{\includegraphics[width=0.3\textwidth]{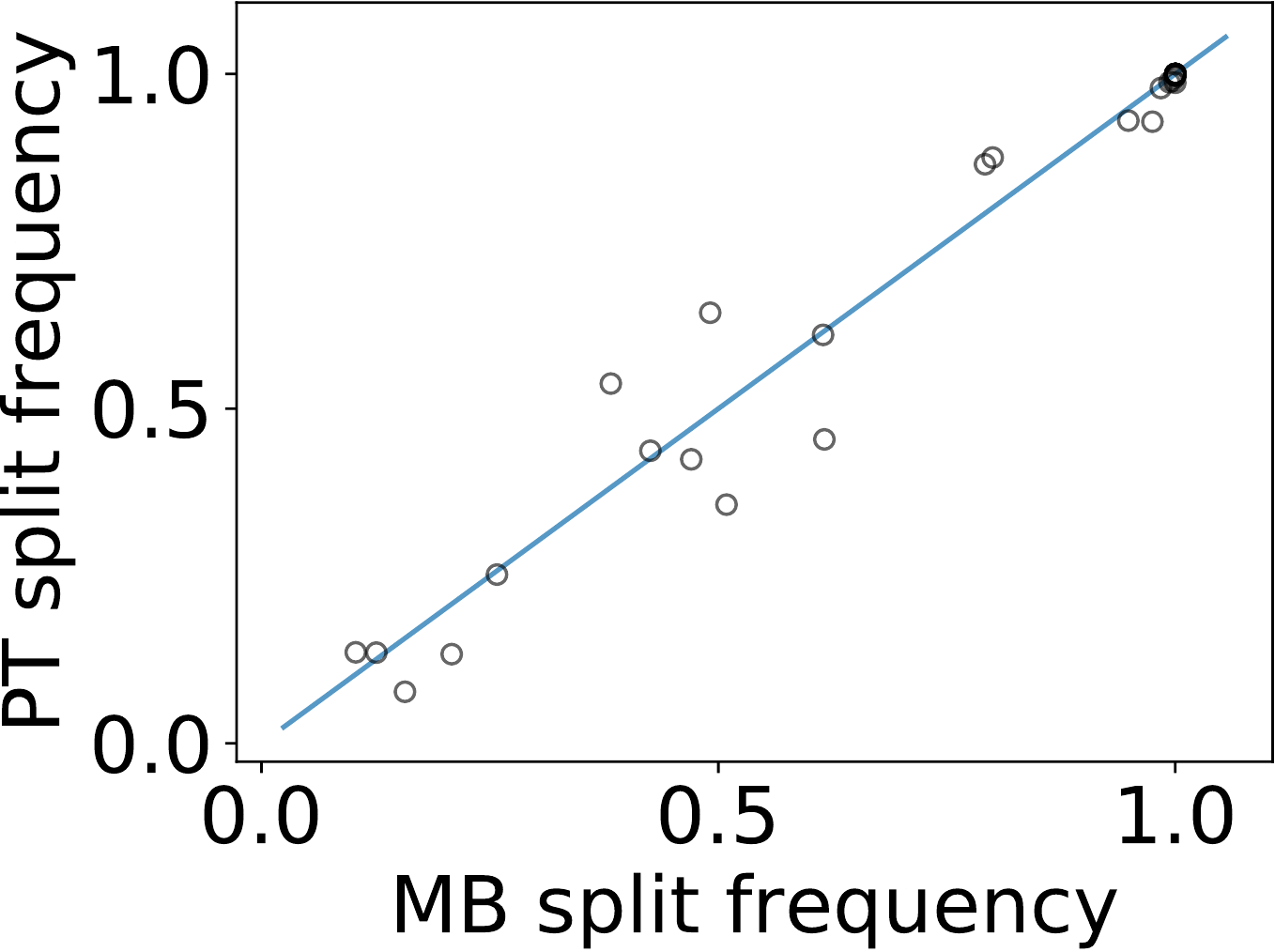}}
	\subfigure[DS3\label{fig:split_scatter_ds3}]{\includegraphics[width=0.3\textwidth]{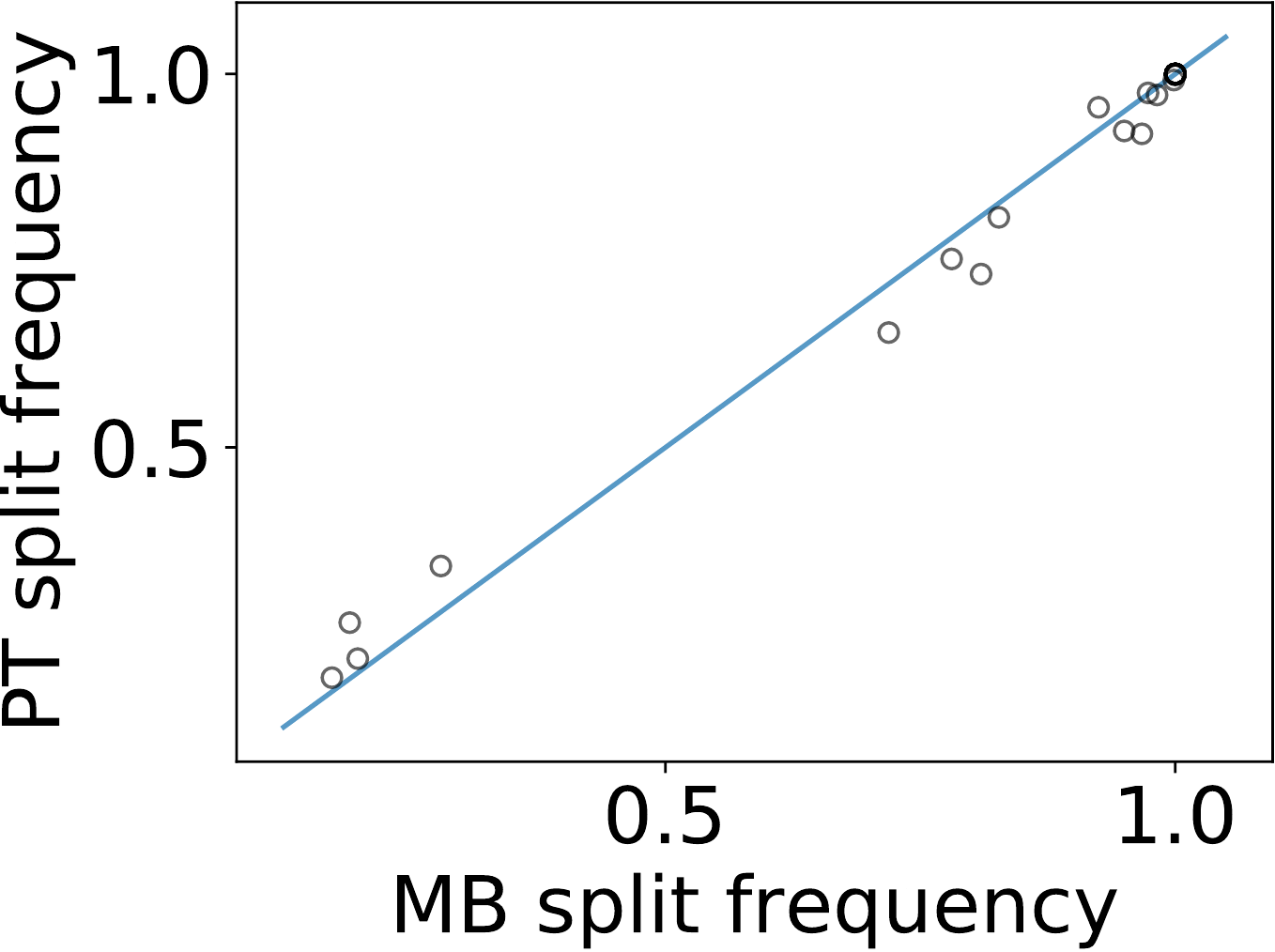}}
	\subfigure[DS5\label{fig:split_scatter_ds5}]{\includegraphics[width=0.3\textwidth]{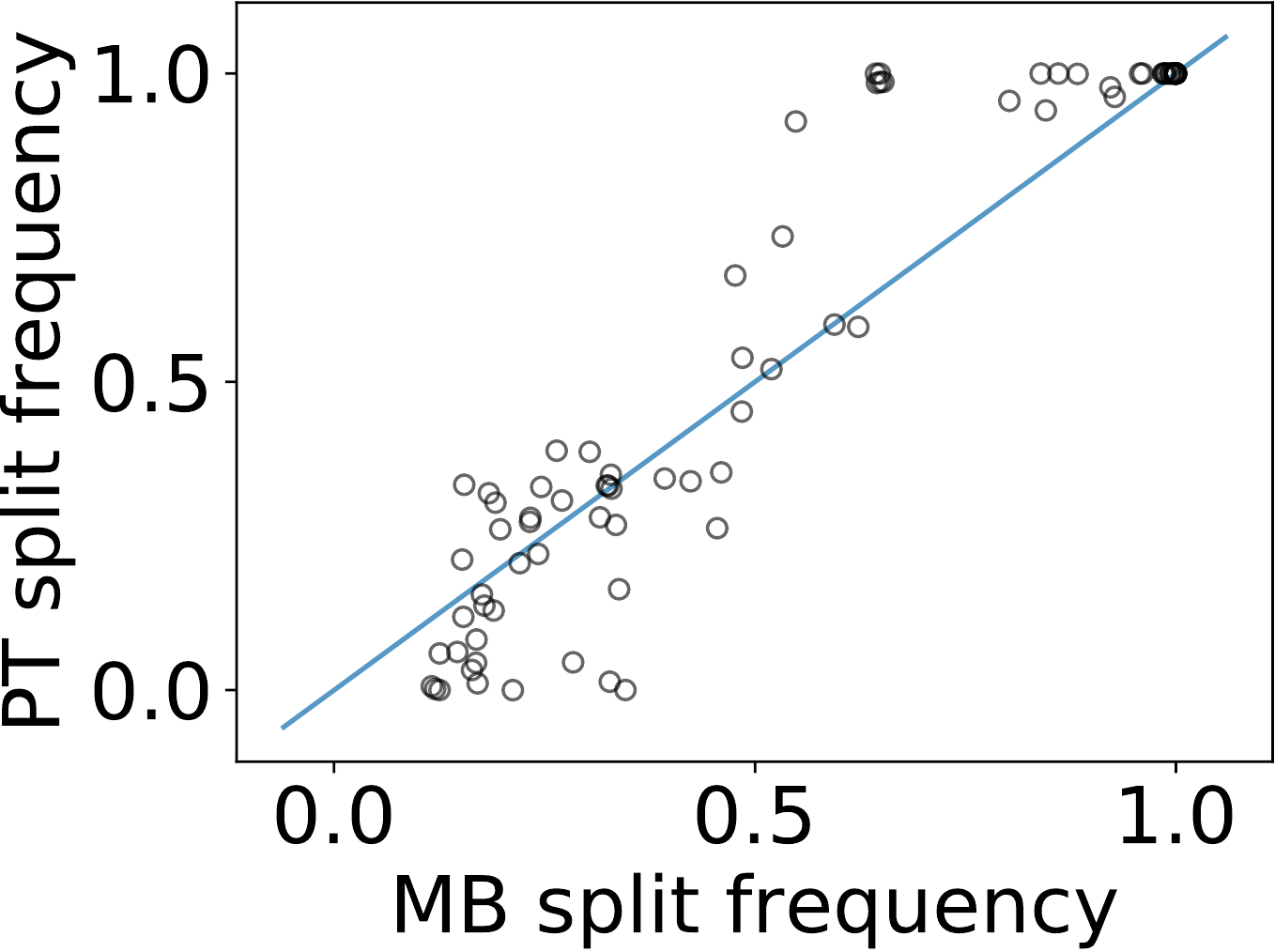}}
	\subfigure[DS7\label{fig:split_scatter_ds7}]{\includegraphics[width=0.3\textwidth]{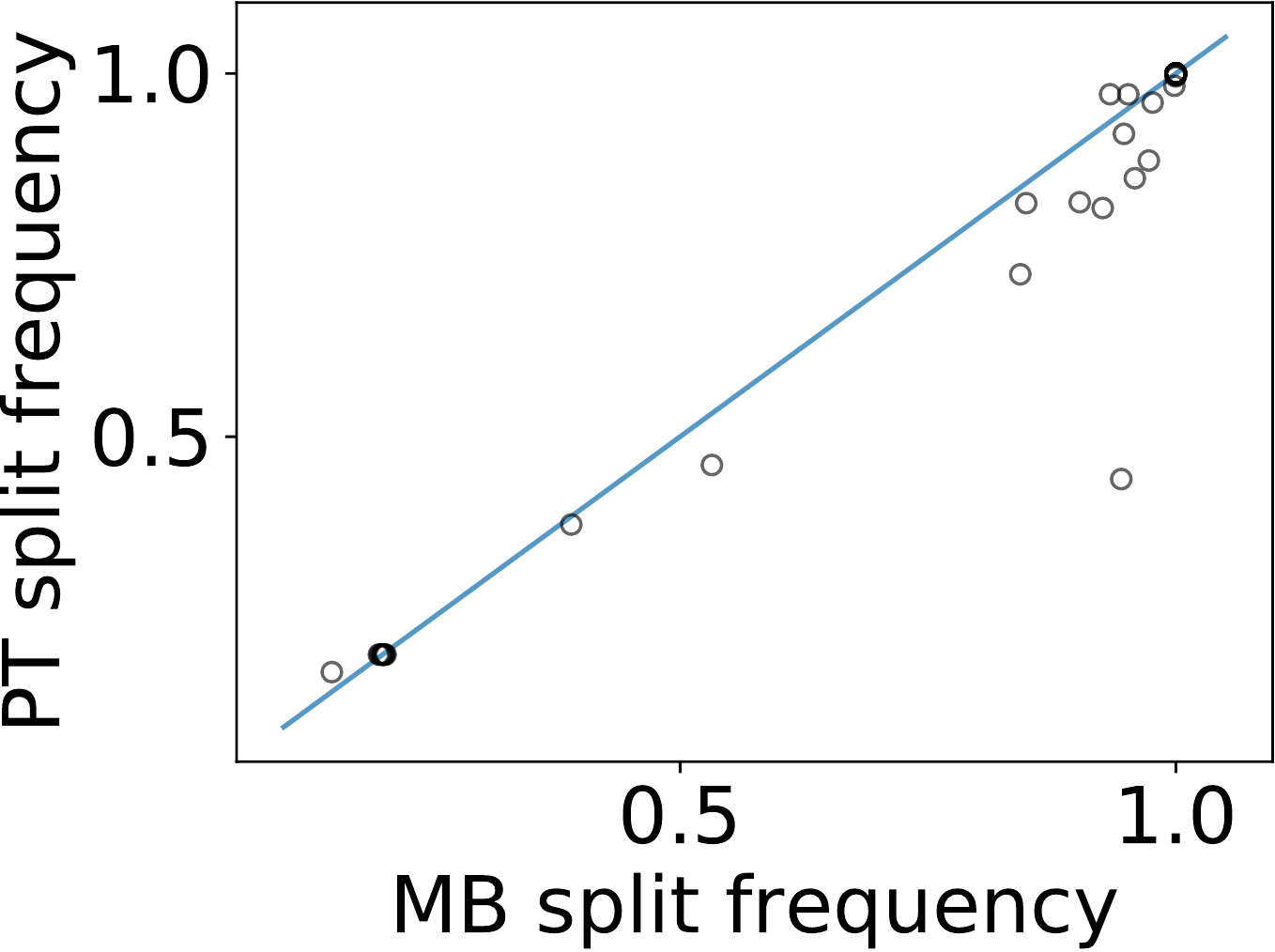}}
	\subfigure[DS8\label{fig:split_scatter_ds8}]{\includegraphics[width=0.3\textwidth]{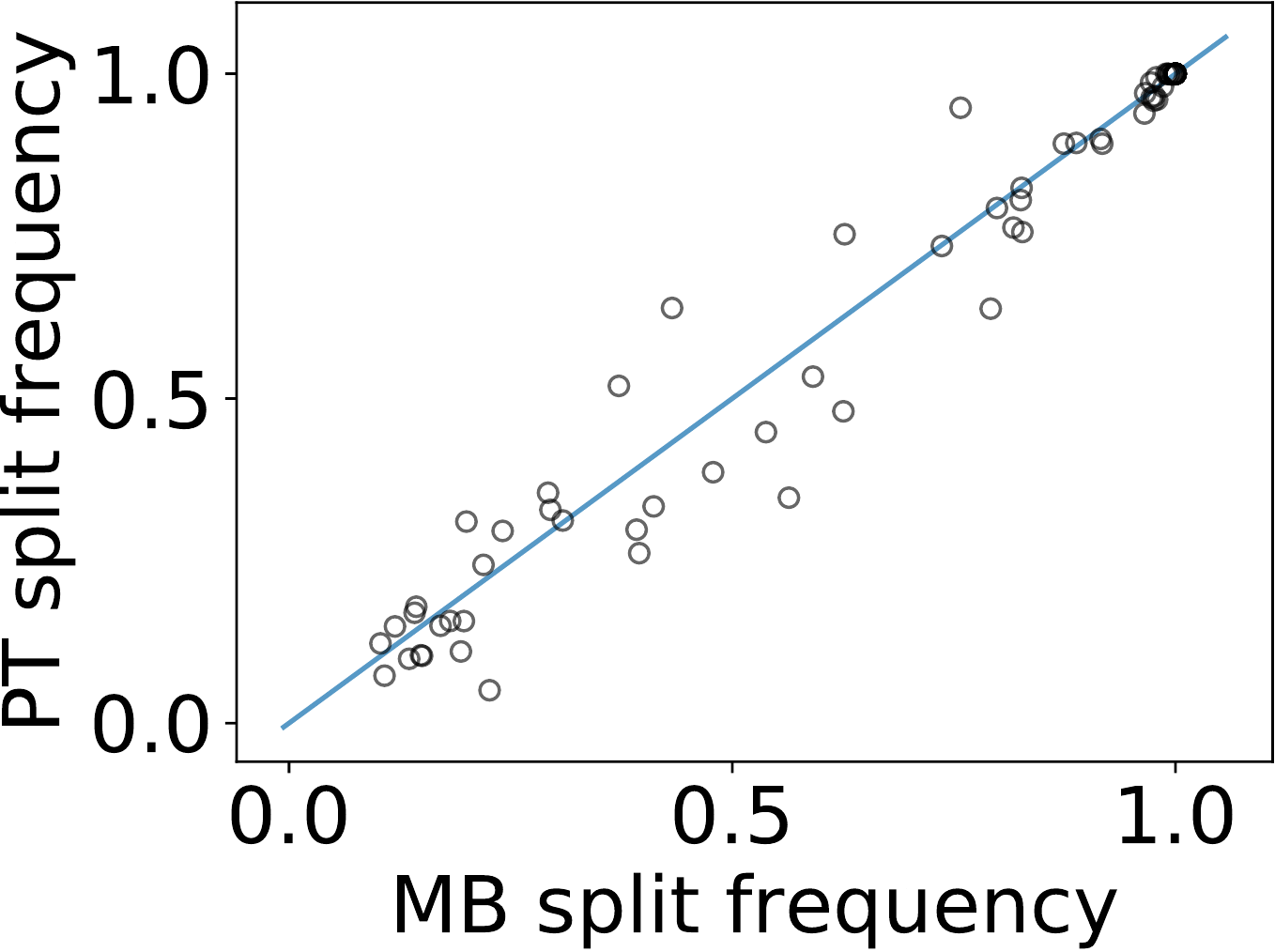}}
	\subfigure[DS10\label{fig:split_scatter_ds10}]{\includegraphics[width=0.3\textwidth]{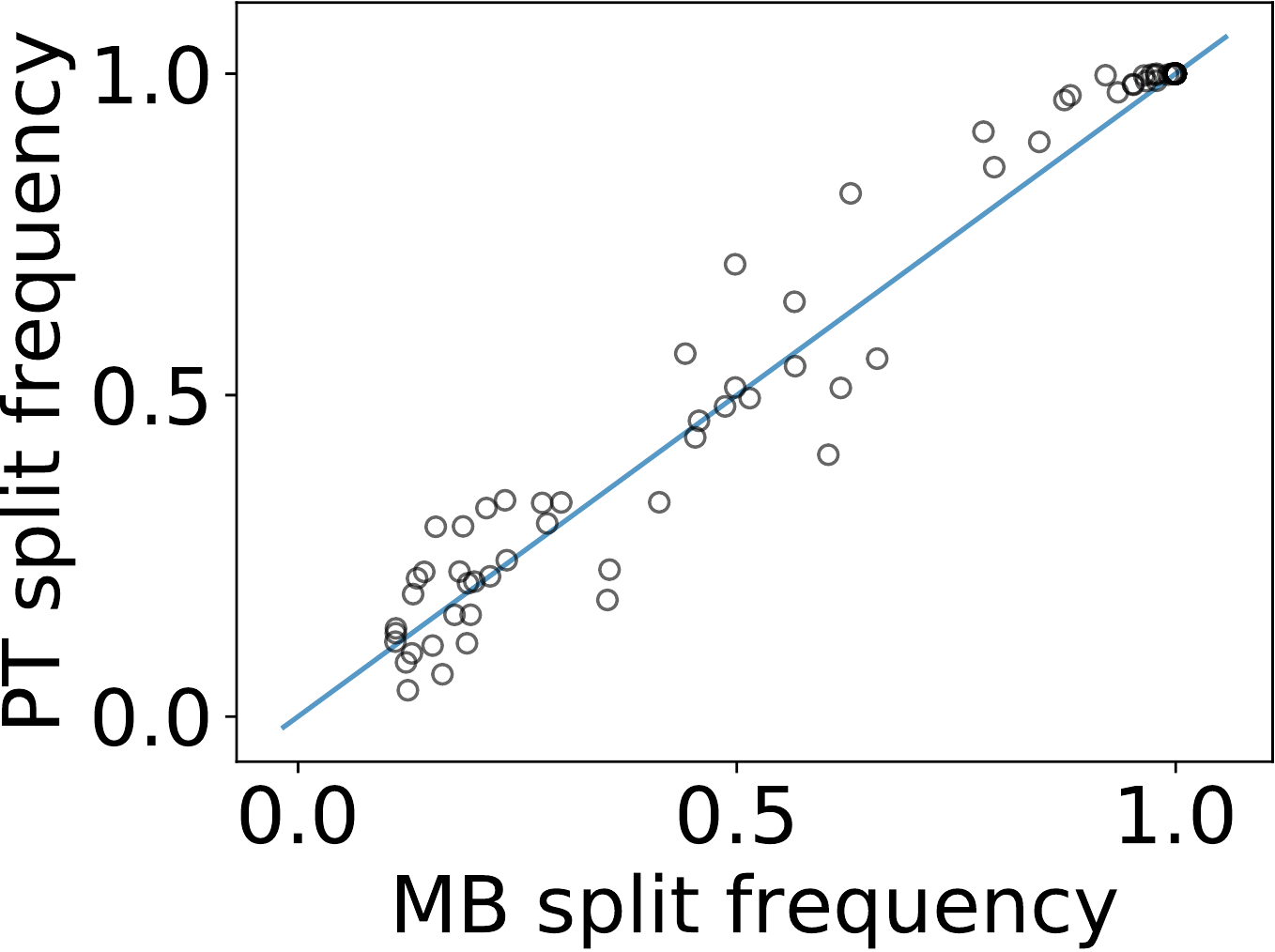}}
	\caption{Comparison of split frequencies from aggregated golden runs over time for MrBayes and PT.}
	\label{fig:split_scatter_supp}
\end{figure}

Split frequencies estimated using maximized likelihood and the HLS were generally similar to the golden run split frequencies for most of our tested datasets (Fig.~\ref{fig:split_scatter}, online supplemental Fig.~\ref{fig:split_scatter_supp}).
However, they did not always agree: one nearly ubiquitous two-taxon split was estimated with a frequency of less than 50\% in DS7, possibly a result of that dataset's lattice-like structure.
Several splits with $> 50\% $ frequency in DS5 were estimated as having very high probability, possibly indicating a failure to find topologies with conflicting splits or a difference in the branch length probability landscape of those trees lost with a simple maximized likelihood estimate.

\begin{table}
\centering
\small
\caption{Differences between consensus topologies generated using split frequencies from MrBayes golden runs and PT HLS with marginal probabilities approximated by maximized likelihood.
We recorded the SPR distance (SPR), RF distance (RF), MrBayes resolved edges (MB-Res), and PT resolved edges (PT-res).
The resolved edge columns count the unique compatible edges of each topology (due to multifurcations in the opposing topology).}
\begin{filecontents*}{data/consensus_trees_distance.csv}
Dataset,SPR, RF, MB-Res, PT-Res
DS1,1,2,0,0
DS2,2,4,0,0
DS3,0,0,0,0
DS4,0,0,1,1
DS5,0,0,0,2
DS6,0,0,1,0
DS7,1,2,1,0
DS8,2,4,1,0
DS10,0,0,1,4
\end{filecontents*}
\csvautotabular{data/consensus_trees_distance.csv}
\label{table:consensus_trees_distance}
\end{table}

\begin{figure}
	\centering
\includegraphics[width=\textwidth]{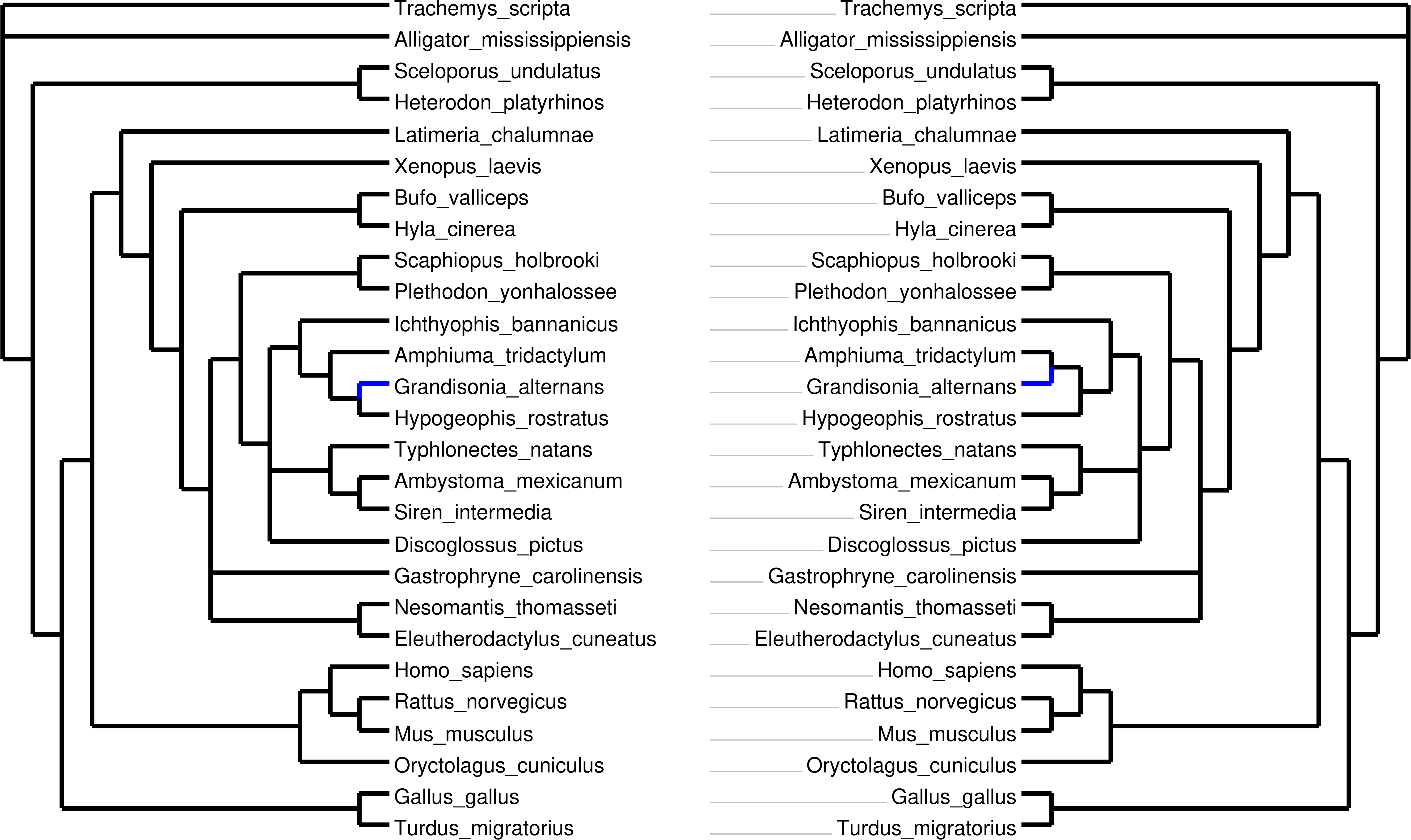}
\caption{A tanglegram showing differences between MrBayes (left) and PT (right) Consensus topologies for DS1.}
\label{fig:consensus_trees_ds1}
\end{figure}

Consensus topologies estimated using maximized likelihood and the HLS were also generally similar to those estimated using the golden run split frequencies (Table~\ref{table:consensus_trees_distance}).
Four of the 10 datasets showed slight disagreement (SPR distance $\le 2$ and RF distance $\le 4$) of at most two edges per topology with poor support (frequency $\le 0.75$ using both methods).
Six of the 10 datasets resolved a small number of edges which remained unresolved in the other topology.
Figure~\ref{fig:consensus_trees_ds1} shows an example of the differences for DS1 generated using Dendroscope 3.5.9~\citep{huson2012dendroscope}.



%
%
%

\section*{Discussion}
We have shown that systematic exploration of phylogenetic tree space using maximum likelihood is a viable strategy for approximating Bayesian posterior distributions.
For the data sets examined, simple maximized likelihood point estimates of topology probability were a quick and useful estimate of marginal topology probability.
The majority of the topologies in the 95\% credible sets identified by MCMC sampling with MrBayes were reachable by starting with a relatively small number of high likelihood topologies and exploring all topologies within up to 10 log-likelihood units.
On datasets with a large number of poorly sampled topologies, PT could not find many of the singleton topologies but did find most of the topologies sampled at least twice by MrBayes.
Moreover, the high likelihood sets identified by phylogenetic topographer had similar split frequency distributions to long MrBayes golden runs, obtaining an RMSD less than 0.05 on all but our flattest dataset.
We tested several search strategies.
Using multiple randomized starting points was a useful strategy on multi-modal datasets which also worked well on single-mode datasets, with little speed penalty.
We observed a good parallelism speed-up of 4.46x with 8 cores using distributed data structures and a work-stealing strategy.
Single branch length optimization of candidate trees followed by full branch optimization of the most promising trees was much quicker than always using full branch optimization with little risk of missing high likelihood topologies.
However, neither applying a prior and estimating MAP values or optimizing the gamma parameter for each visited topology proved useful for finding high likelihood topologies and the latter came with a substantial performance tradeoff.
These strategies may prove more useful under more complex phylogenetic models and are worth further study.
It may also be interesting to try other tree rearrangements, such as SPR.
Searching with SPRs may allow using a smaller negative threshold $T$ because PT could jump between peaks without going through low likelihood topologies.
The tradeoff is that employing SPR rearrangements would slow down computation due to a substantial increase in the number of neighbors, so practical usage of SPR with PT may require novel algorithms or optimization.

We believe that the HLS identified by PT will be useful for a number of applications.
The most direct application, to obtain approximate Bayesian posterior distributions, will be enabled by better estimates of the marginal probability of HLS tree topologies.
We have recently developed efficient such methods \citep{dubious} that we show work well for common models, and hope to show that they work well for more complex phylogenetic models in the future.
One could also develop proposal distributions for MCMC samplers which draw topologies directly from the HLS set \citep{Pollard2018-xz}, or use an importance-sampling approach.
Another related application is to rapidly obtain a consensus tree along with approximate split frequency estimates.
In our tested datasets, the HLS explored by PT provided a useful estimate of the split frequencies in minutes and the estimated consensus topologies differed only slightly from those estimated using long MrBayes golden runs.

Although the PT strategy as we have formulated it is novel, it shares goals and strategies with previous work.
This line of work begins with \citet{Maddison1991-tc}, who had the goal of finding ``islands'' of most-parsimonious topologies using hill-climbing by using many starting points.
However, this work focused on finding these islands, rather than collecting all topologies that have a given level of optimality around them as we do here.
\citet{Salter2001-ar} extended this work from parsimony to likelihood, while others \citep{Salter2001-wa,Vos2003-xq,Zwickl2006-zx,Nguyen2015-bs} have developed strategies to find maximum-likelihood trees in the presence of multiple likelihood peaks.
Our work is distinguished from this previous work by our goal of approximating tree topology posterior distributions using likelihood methods, and by our methods of parallelized systematic exploration of suboptimal topologies.
Another vein of work uses a collection of trees obtained using bootstrapping as a proxy for a posterior distribution \citep{Rodrigo2009-hr,Syme2012-hu,Pankey2014-go,Pollard2018-xz}.
That work uses different methods and has not shown the level of agreement that we obtain between our approximation and the true posterior.

The PT approach also has precedent in other areas of statistics.
One connection is with the idea of Bayesian model averaging \citep{Hoeting1999-qh}, in which here the model to be averaged over is the phylogenetic tree topology.
If we restrict parameter estimation in Bayesian phylogenetics to the PT set, this becomes the ``Occam's window'' strategy of \citet{Madigan1994-mw}.
Furthermore, \citet{Dobra2010-tb} propose MOSS, a local search strategy over log-linear models with conjugate priors to find the set in the Occam's window; like PT it also records models to avoid revisiting them.
However, in our setting, we do not have easy access to marginal likelihoods through conjugacy.
Also, MOSS is a stochastic search procedure to avoid being stuck in local optima, whereas we identify local optima ahead of time using hill-climbing and have a search that does not include any explicit stochasticity in the design.

PT is part of a larger trend in Bayesian statistics away from sampling-based methods such as MCMC and towards optimization-based methods such as variational Bayes \citep{Blei2017-fs}.
Here we find an approximate posterior distribution on a discrete collection of items using direct search.
Much remains to be done to build on this starting point.
Our current search strategy is not smart-- it simply tries every NNI at every location and must consider paths of equivalent moves such as moving a taxon $A$ and then a taxon $B$ or moving $B$ first and then $A$.
Also, to be viable in the general case, PT needs to overcome the challenge posed by very diffuse posteriors; because such diffuse posteriors appear to result from unresolved splits~\citep{Whidden2015-eq} we may look to collapse ambiguous splits or express topology posteriors in a factored form \citep{larget2013estimation,Zhang2018-ld}.
More advanced techniques and search strategies such as deep reinforcement learning \citep{Mnih2015-so,Silver2016-cy}, combined with better marginal likelihood estimates, may form the basis for fast new Bayesian phylogenetic methods.

\section*{Funding}
This research was funded by National Science Foundation grant CISE-1564137 and National Institutes of Health grant U54-GM111274.
The research of CW was supported as a Simons Foundation Fellow of the Life Sciences Research Foundation.
The research of FAM was supported in part by a Faculty Scholar grant from the Howard Hughes Medical Institute and the Simons Foundation.

\section*{Acknowledgments}
We wish to thank summer interns Lola Bradford and Andrew Wei for their work implementing and testing local maxima estimation as part of a related project (in preparation).
We also wish to thank Arman Bilge, Vu Dinh, and Vladimir Minin for helpful comments and suggestions.

\bibliographystyle{sysbio}
\bibliography{pt}

\clearpage

\notarxiv{\markboth{}{}}
\beginsupplement
\clearpage
\section{Supplemental Figures}

\end{document}